\documentclass[11pt,a4paper]{article} % Plantilla neutra ideal para arXiv

\usepackage[scaled]{helvet} % Carga Helvetica
\usepackage[utf8]{inputenc}
\usepackage[margin=1in]{geometry}

\usepackage{graphicx}%
\usepackage{multirow}%
\usepackage{amsmath,amssymb,amsfonts}%
\usepackage{amsthm}%
\usepackage{mathrsfs}%
\usepackage[title]{appendix}%
\usepackage{xcolor}%
\usepackage{textcomp}%
\usepackage{manyfoot}%
\usepackage{booktabs}%
\usepackage{algorithm}%
\usepackage{algorithmicx}%
\usepackage{algpseudocode}%
\usepackage{listings}%
\usepackage{float}
\usepackage{graphicx}
\usepackage{subcaption}
\usepackage{placeins}
\usepackage{longtable}

\usepackage{hyperref}
\usepackage{xurl}
\hypersetup{breaklinks=true}

\usepackage{authblk} % Paquete estándar para múltiples autores y afiliaciones

\title{Soroll-IA: A Weakly Labeled Audio Dataset for Real-World Industrial Port Monitoring}

\author{Javier Naranjo-Alcazar}
\author{Jordi Grau-Haro}
\author{Ruben Ribes-Serrano}
\author{Marta Garcia-Ballesteros}
\author{Pedro Zuccarello}

\affil{Instituto Tecnologico de Informatica (ITI), Valencia, Spain}

\date{} % Deja esto vacío si no quieres que aparezca la fecha actual automáticamente

\begin{document}

\maketitle

\abstract{Soroll-IA is a weakly labeled environmental audio dataset recorded in a real-world industrial port environment in Valencia (Spain) using two fixed sensing nodes. The dataset comprises approximately 22 hours of audio segmented into 7,396 clips and covers 26 sound event classes representative of industrial port acoustic activity commonly observed in such environments, such as crane sirens, train movements, traffic, and other logistical and industrial sounds. Recordings were captured under highly challenging acoustic conditions, including strong background noise, long-distance sources, and frequent event overlap. All audio clips were annotated by domain experts following a weak labeling strategy, where tags indicate the presence of sound events within a clip without temporal localization. To account for inter-annotator variability, two ground-truth versions are released: one without cross-validation, where a class is considered present if annotated by at least one expert, and a second, more conservative version based on cross-validation, where agreement by at least two-thirds of the annotators is required. The dataset is intended to support research in audio tagging, weakly supervised sound event detection, and machine learning under realistic industrial acoustic conditions. Benchmark results are provided using two complementary architectures: CNN14 from the PANNs family, representing high-capacity convolutional models for audio tagging, and MobileNetV2, selected for its suitability in real-time classification on low-resource edge devices. To the best of current knowledge, Soroll-IA constitutes an available dataset dedicated exclusively to industrial port acoustic environments, aiming to foster advances in robust environmental sound analysis for safety-critical and operational monitoring applications. The dataset is available online\footnote{\url{https://www.kaggle.com/datasets/itiresearch/soroll-ia-weakly-labeled-audio-port-monitoring/}} \hspace{0.01pt} and collected under Attribution-NonCommercial 4.0 International license.}


\section{Introduction}\label{sec:intro}

\begin{figure}
    \centering
    \includegraphics[width=0.75\linewidth]{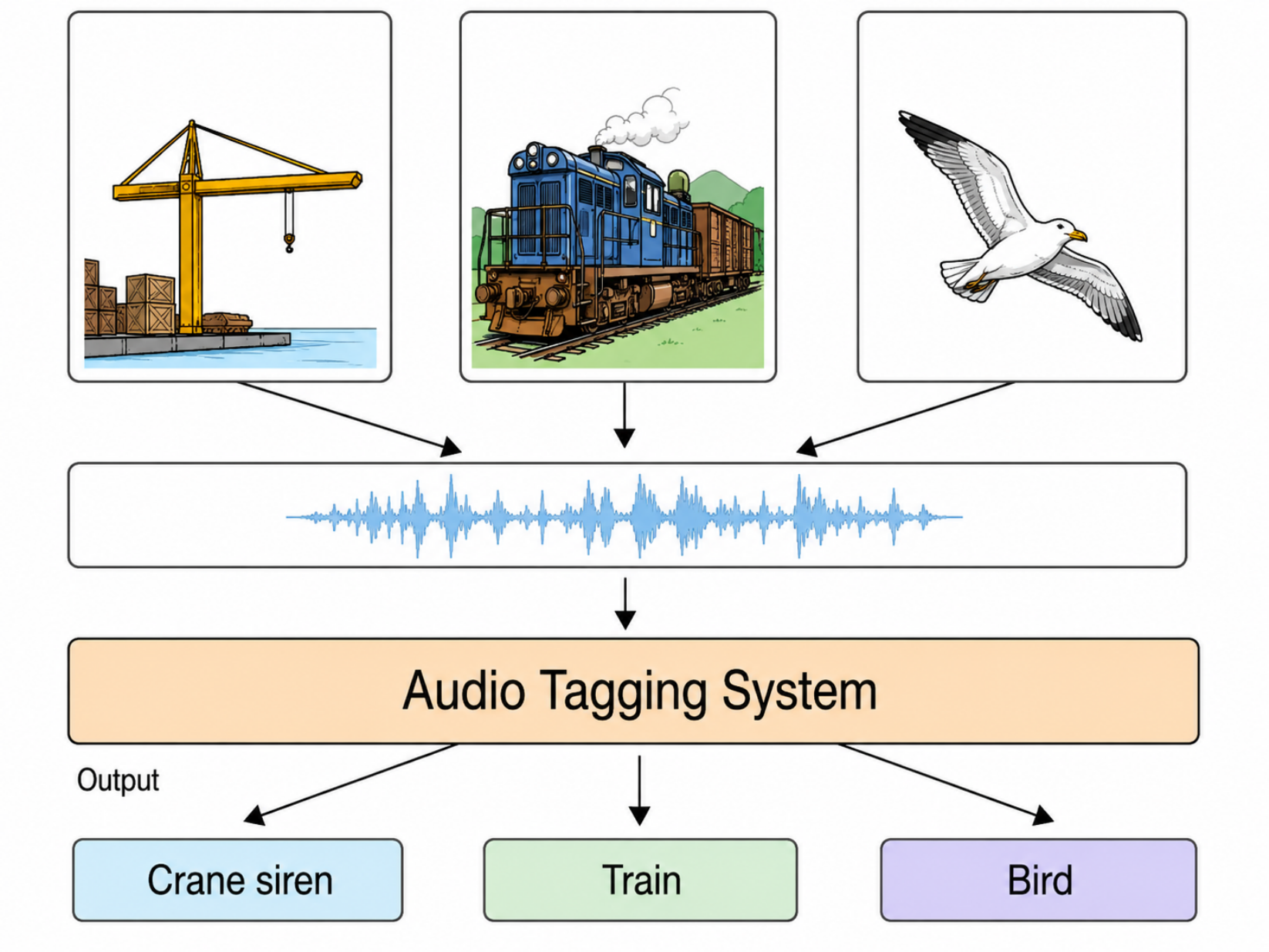}
    \caption{Representation of an Audio Tagging system in the context of the Soroll-IA dataset}
    \label{fig:audio_tagging}
\end{figure}

Audio tagging—the task of assigning one or multiple labels to an audio segment to indicate the presence of specific sound events—is a fundamental problem in computer audition \cite{triantafyllopoulos2025computer, gong2021psla, singh2024atgnn, schmid2023efficient} (see Fig.~\ref{fig:audio_tagging}). Its relevance stems from enabling scalable indexing, retrieval, and analysis of large audio collections without requiring precise temporal annotations, thereby substantially reducing labeling costs compared to strongly supervised approaches such as Sound Event Detection (SED) \cite{mesaros2021sound, schmid2025effective}. However, the effectiveness of this paradigm critically depends on the quality of the weak labels, as their inherent imprecision can negatively impact model performance \cite{annotation}. To mitigate this issue, samples must be annotated by human experts, which, although increasing the overall effort and cost of dataset creation, ensures more reliable labels, provides valuable insights into the data, and facilitates the identification and systematic organization of acoustic events into well-defined classes.

Despite the increase in audio tagging solutions in the state of the art, available public datasets remain limited in scope with respect to challenging outdoor and industrial sound environments. The majority of existing corpora focus on either urban street scenes (e.g., AudioSet \cite{gemmeke2017audio}, FSD50K \cite{fonseca2021fsd50k}), indoor machine operations (e.g., MIMII \cite{Purohit2019}), or domestic acoustic scenes (e.g., ESC-50 \cite{piczak2015esc} or Chime-Home \cite{chime}), and typically lack labeling aligned with port or industrial noise contexts.

To address this gap, Soroll-IA has been curated as a weakly labeled, multilabel audio dataset recorded via fixed outdoor sensor nodes in one demanding real-world environment: industrial ports. The dataset comprises approximately 22 hours of audio, divided into 7,396 clips, each with annotations for potentially overlapping events such as crane sirens, train noise, heavy traffic, reversing beeps, and drill sounds, but without timestamped boundaries. This structure supports realistic modeling of ambient event tagging under noisy, dynamic conditions.

Releasing Soroll-IA is meaningful for several reasons. First, it provides a domain-specific resource that fills a gap between urban acoustic datasets and indoor industrial corpora, allowing models to be tested on ambient, industrially relevant, outdoor audio. Second, the dataset’s weak multilabel format reflects real-world annotation constraints and encourages development of algorithms capable of managing label noise and ambiguity—challenges echoing broader weak supervision research. Finally, supplying a reproducible and openly available benchmark fosters progress and comparability in environmental sound analysis—particularly in scenarios critical for safety, infrastructure monitoring, and smart industry.

In summary, Soroll-IA represents one of the first publicly available collections of real-world port ambient audio, equipped with weak, multilabel annotations and captured under outdoor noise-rich conditions. It thus offers a timely and valuable resource for advancing audio tagging techniques tailored to challenging acoustic domains.

The remainder of this paper is organized as follows. Section~\ref{sec:related} reviews the most widely used datasets in the computer audition community. Section~\ref{sec:data} describes the technical aspects involved in the creation of the Soroll-IA dataset, including the recording hardware and deployment locations, the labeling strategy, and the post-processing steps applied after annotation. Section~\ref{sec:sorollia} details the released data and its structure. Section~\ref{sec:baseline} presents benchmark results that serve as a reference baseline for future research. Finally, Section~\ref{sec:conclusion} concludes the paper and outlines potential directions for future work.

\section{Related Work}\label{sec:related}

\subsection{State of the Art Datasets}\label{subsec:datasets}

Several publicly available audio datasets address related tasks such as environmental sound classification, urban noise monitoring, or machine condition analysis. However, none fully align with the scope and characteristics of Soroll-IA, which focuses exclusively on real-world outdoor industrial port acoustic environments and provides weak, clip-level multilabel annotations. In contrast to existing datasets, Soroll-IA captures port-specific sound events—including crane sirens, train movements, traffic, and other logistical activities—recorded by fixed outdoor sensing nodes under highly variable and noisy conditions. This combination of domain specificity, weak labeling, and long-term ambient recordings distinguishes Soroll-IA from prior datasets and addresses a gap in publicly available resources for industrial port audio analysis.

Several publicly available datasets address complementary aspects of environmental and industrial audio analysis, yet none fully match the scope and characteristics of Soroll-IA. For instance, the MAVD-traffic dataset \cite{Zinemanas2019} provides outdoor audio–video recordings for sound event detection in urban environments, featuring a hierarchical ontology that combines vehicle categories and mechanical components with associated actions. While MAVD-traffic is richly annotated and captures real-world traffic dynamics, its focus on urban road scenarios and traffic-centric events limits its representativeness of industrial port acoustics, where sound sources, spatial layouts, and operational activities differ substantially.

Another well-established line of work focuses on industrial machine sounds, exemplified by datasets such as MIMII and MIMII-DUE \cite{Purohit2019, mimiidue}, which target anomaly detection in machinery including valves, pumps, fans, gearboxes, and slide rails. These datasets provide high-quality recordings under controlled operating conditions and typically rely on strong labels to characterize normal and anomalous behavior. However, they are predominantly recorded in indoor or semi-controlled environments and emphasize machine condition monitoring rather than the complex, overlapping soundscapes observed in outdoor industrial infrastructures.

Urban traffic-oriented datasets such as IDMT-Traffic \cite{abesser2021idmt} collect strongly labeled vehicle sounds, often annotated by vehicle type or motion direction. While useful for classification tasks, these datasets usually focus on isolated events and lack the multilabel, polyphonic characteristics that arise naturally in continuous ambient recordings from operational industrial sites. Similarly, UrbanSound8K \cite{urbansound8k} provides outdoor urban audio clips covering common sound classes such as sirens, drilling, or engine noise. Despite its widespread adoption, UrbanSound8K is composed of short, single-label excerpts extracted from urban contexts and does not capture the sustained ambient conditions, overlapping events, or domain-specific activities characteristic of industrial port environments.

At the opposite end of the spectrum, large-scale corpora such as AudioSet \cite{gemmeke2017audio} and FSD50K \cite{fonseca2021fsd50k} offer weakly labeled audio clips spanning hundreds of sound classes and are frequently used as general-purpose training datasets. In particular, FSD50K relies heavily on crowdsourced annotations and web-derived recordings, resulting in broad coverage but limited control over recording conditions and label consistency. In contrast, Soroll-IA is built from in situ recordings obtained via fixed outdoor sensors in an operational industrial port and is manually annotated by domain experts, prioritizing label reliability and contextual relevance over dataset scale.

Scene-centric datasets, including TAU Urban Acoustic Scenes and its MACS/MATS extensions \cite{Mesaros2018_DCASE, Heittola2020}, contribute valuable resources for acoustic scene classification and audio captioning. Nevertheless, their scene-level focus and limited representation of industrial port events constrain their applicability to event-level audio tagging in industrial monitoring scenarios.

In summary, Soroll-IA distinguishes itself through its exclusive focus on real-world outdoor industrial port environments, the use of continuous ambient recordings acquired by fixed sensing nodes, and the provision of weak, multilabel annotations tailored to port-specific acoustic events such as crane sirens, train movements, and traffic activity. This unique combination positions Soroll-IA at the intersection of environmental sound analysis, industrial monitoring, and weakly supervised learning, addressing a gap not covered by existing public datasets.

\subsection{Motivation}\label{subsec:motivation}

Traditional Internet of Things (IoT) frameworks commonly rely on centralized cloud-based processing, which introduces latency and limits the timely detection of critical events in dynamic and safety-sensitive environments \cite{informatics11040071}. In industrial contexts such as seaports—where alarms, machinery malfunctions, and logistical activities must be detected promptly—delayed analysis can significantly undermine operational efficiency and situational awareness. Despite their economic and strategic importance, industrial ports remain largely unexplored within the computer audition literature, particularly when compared to urban traffic or indoor industrial monitoring scenarios \cite{bello2019sonyc, ooi2021strongly}. This gap is notable, as real-time acoustic monitoring in ports could support applications such as the detection of abnormal events, identification of logistical delays, and continuous assessment of noise pollution, all of which have direct implications for port safety and operational optimization.

Deploying computer audition inference capabilities directly at the edge addresses these limitations by enabling continuous, low-latency audio tagging from fixed outdoor recording nodes. Edge-based processing reduces dependence on network connectivity, lowers bandwidth requirements by avoiding the transmission of raw audio streams, and enhances data privacy \cite{bello2019sonyc}. Furthermore, it enables scalable monitoring of complex and highly variable soundscapes characteristic of industrial ports, where multiple sound sources—such as cranes, trains, and vehicular traffic—often overlap in time and space. These requirements motivate the need for robust audio tagging models trained on domain-specific data that accurately reflect real-world port acoustics.

The integration of distributed recording nodes with IoT platforms further strengthens this paradigm by enabling centralized visualization, remote supervision, and scalable management of large sensor networks. IoT-based dashboards allow operators to monitor acoustic activity across multiple locations in real time and to extract actionable insights from audio tagging outputs. For instance, remote monitoring frameworks based on platforms such as ThingsBoard have been successfully employed in port environments to supervise infrastructure status and environmental conditions, facilitating informed decision-making and operational control \cite{Naranjo:2025:AES}. The real-world deployment of recording nodes at the Port of Valencia (Spain) exemplifies this approach, highlighting both the feasibility and the practical relevance of continuous acoustic monitoring in operational port settings. This context directly motivates the creation of Soroll-IA, a dataset designed to advance research in audio tagging for outdoor industrial port environments and to support the development of efficient, edge-oriented computer audition solutions.

\section{Data Collection and Labeling}\label{sec:data}

\begin{figure}
    \centering
    \includegraphics[width=0.65\linewidth]{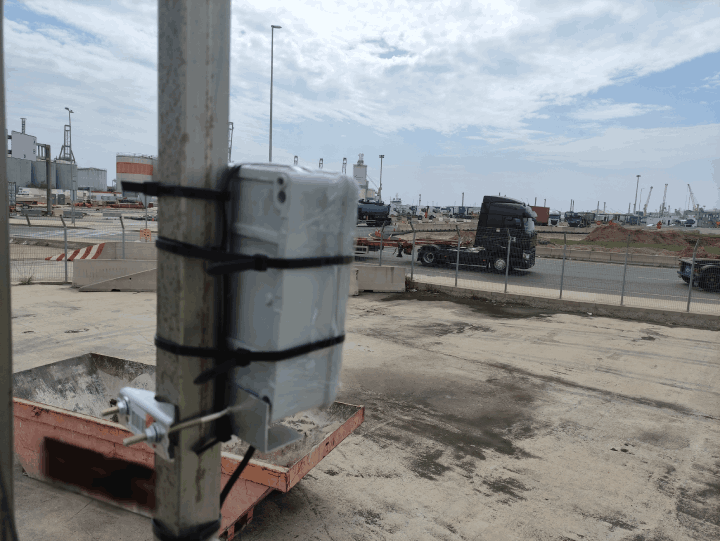}
    \caption{Example of a recording node that obtains audio from traffic.}
    \label{fig:node1}
\end{figure}

\begin{figure*}[t]
    \centering
    \begin{subfigure}{0.48\linewidth}
        \centering
        \includegraphics[width=\linewidth,keepaspectratio]{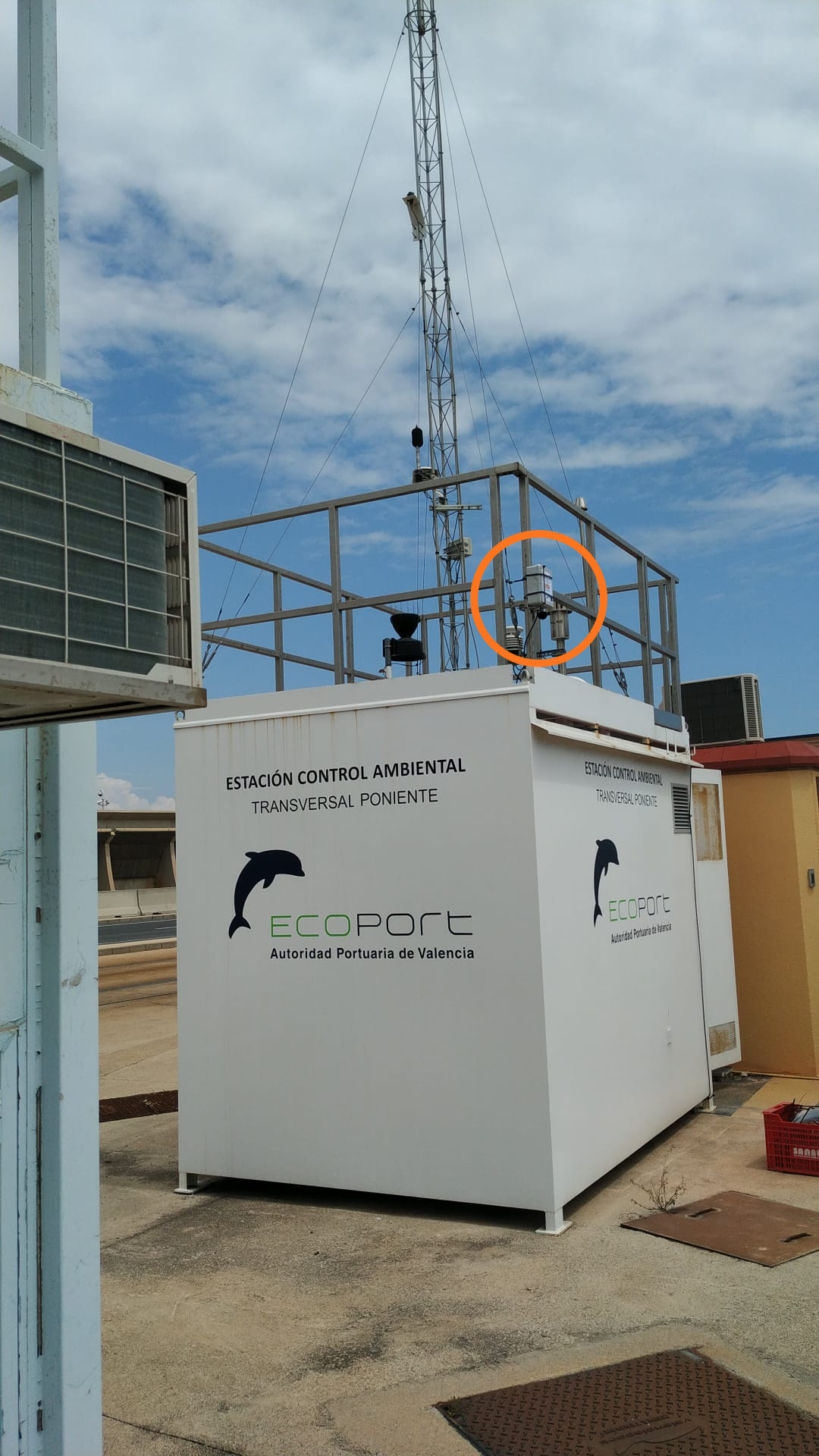}
        \caption{Recording node deployed near the train rails}
    \end{subfigure}
    \hfill
    \begin{subfigure}{0.48\linewidth}
        \centering
        \includegraphics[width=\linewidth,keepaspectratio]{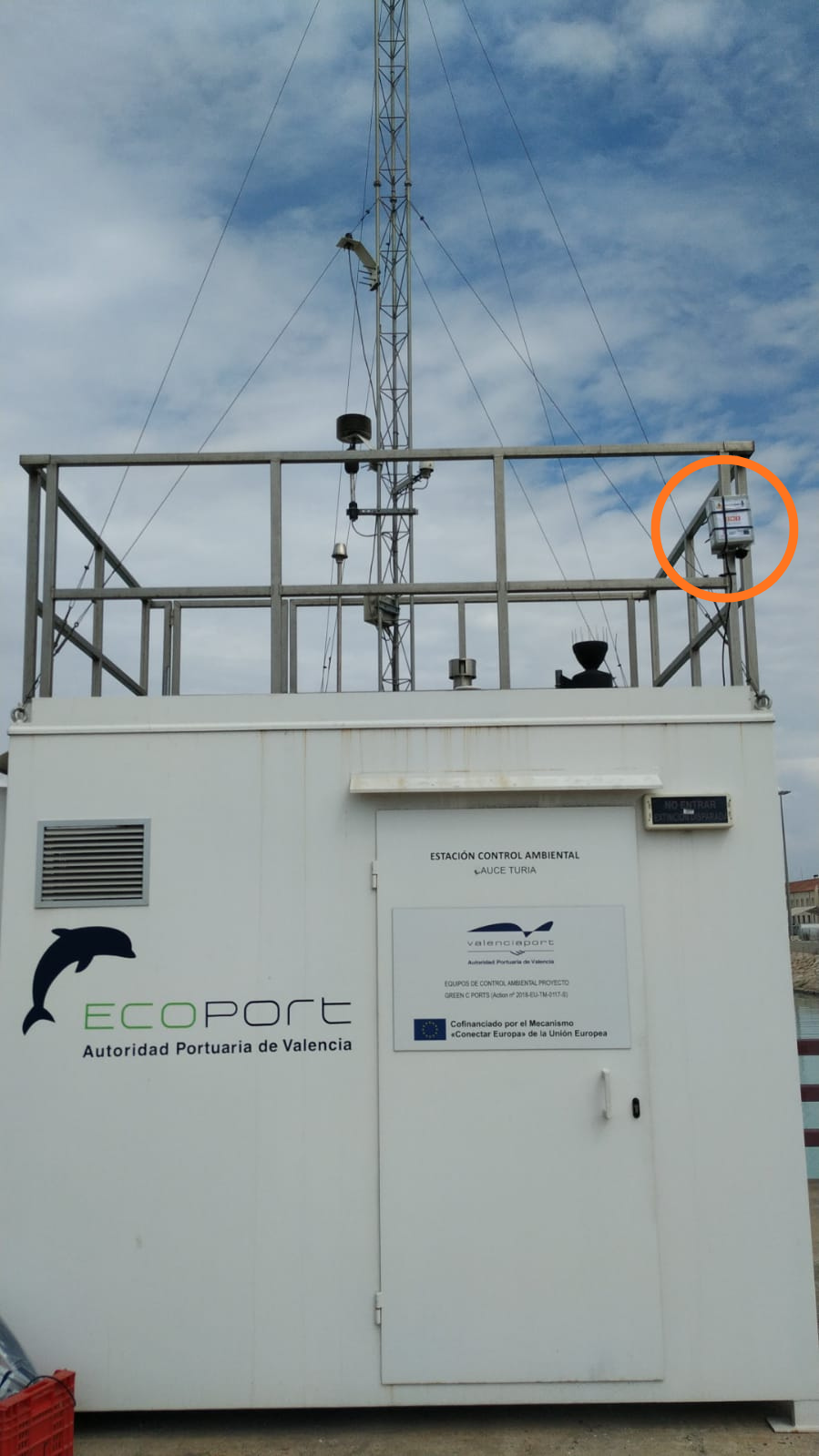}
        \caption{Recording node deployed near a roundabout}
    \end{subfigure}
    \caption{Environmental control cabins where the two recording nodes are located. Recording nodes are marked with orange circles.}
    \label{fig:nodes}
\end{figure*}

This section describes the process followed for obtaining audio recordings and their corresponding annotations, detailing the recording nodes, the characteristics of the collected data, and the labeling procedure.

\subsection{Recording node}\label{subsec:node}

The audio data in Soroll-IA were collected using fixed recording nodes specifically designed for long-term deployment in outdoor industrial environments. These nodes were conceived to operate continuously under real-world conditions, such as ports where weather exposure, background noise variability, and limited maintenance access pose significant challenges.

Each recording node integrates a calibrated omnidirectional microphone (specifically, a Rode Lavalier II microphone connected via a dedicated USB sound card) connected to an embedded processing unit (based on a Raspberry Pi platform) capable of performing scheduled audio acquisition and local preprocessing. The hardware configuration prioritizes robustness, low power consumption, and stable long-term operation, enabling unattended recording over extended periods. To ensure continuous operation under harsh outdoor conditions, the components are housed inside a weather-resistant PVC IP65 enclosure (see Fig.~\ref{fig:node1}). Audio signals are captured at a fixed sampling rate (44.1 kHz) and segmented into short-duration clips suitable for downstream audio tagging and analysis tasks (10-second audio clips).

Beyond raw audio acquisition, the nodes are designed to support edge-oriented workflows, allowing local control of recording schedules and seamless integration with higher-level monitoring platforms. Metadata such as timestamps, node identifiers, and deployment location are systematically associated with each recording, facilitating large-scale dataset management and traceability across distributed sensors.

The design and deployment strategy of these recording nodes follows a data-centric approach, where data acquisition, storage, and annotation constraints are jointly considered from the outset. This infrastructure was originally introduced and validated in a real-world port environment, demonstrating its suitability for industrial acoustic monitoring and dataset construction. A detailed description of the node architecture, deployment strategy, and data management pipeline is provided in our previous work \cite{aes_berlin}, where the recording system was used to support active learning–driven annotation under limited labeling budgets.

By using this validated recording infrastructure, Soroll-IA ensures consistency in data quality while enabling scalable acquisition of complex outdoor soundscapes. This approach supports the creation of datasets that better reflect the acoustic conditions encountered in operational industrial and construction scenarios, bridging the gap between controlled laboratory recordings and real-world audio monitoring applications.

Soroll-IA is composed of audio recordings collected from two fixed recording nodes deployed at strategically distinct locations within the industrial port of Valencia. The first node was installed near the main entrance of the port, located at the top of an environmental monitoring control stand located adjacent to a traffic roundabout (see Fig.~\ref{fig:node1} and Fig.~\ref{fig:nodes}-(b)). During the recording period, road improvement works were carried out in this area, resulting in a soundscape characterized by both continuous traffic activity and construction-related events. Consequently, this node captures a diverse set of acoustic events such as Jackhammer, Drill, Idling, Traffic, among others, often occurring under high background noise and overlapping conditions. The second node was deployed at the top of a control booth located next to the railway tracks, with port cranes positioned behind the installation (see Fig.~\ref{fig:nodes}-(a)). This placement leads to recordings dominated by rail and port logistics activity, including events such as Train Wheels Squealing and Siren, the latter corresponding to warning signals emitted by port cranes, along with other operational industrial sounds.

\subsection{Labeling procedure}\label{subsec:labeling}

\subsubsection{Annotators and labeling protocol}

The annotation process involved five expert annotators, all of whom are data scientists with experience in audio signal processing and computational audio analysis, and were therefore familiar with audio tagging and sound event classification tasks. Two of them were directly involved in the deployment of the acoustic sensing nodes in the port environment and therefore had extensive domain knowledge of the operational context. These annotators can be considered field-experienced experts with strong familiarity with the acoustic characteristics of the monitored environment.

To structure the annotation process, annotators were organized into two groups (three annotators and two annotators, respectively). The two annotators with direct field experience in the deployment of the acoustic sensing nodes were deliberately assigned to different groups to ensure that domain-experienced expertise was present in both annotation teams. Each audio clip was independently annotated within its assigned group, with each clip being reviewed by either three or two annotators, depending on the group structure.

Each annotator processed 40 audio clips per day (corresponding to 200 clips per group per week), where each clip had a duration of approximately 10 seconds, resulting in an estimated annotation effort of around 30 minutes per day per annotator.

Annotations were performed using the BAT\footnote{\url{https://github.com/BlaiMelendezCatalan/BAT}} annotation tool, which was slightly modified to integrate into the active learning workflow.

A dedicated ``Doubt'' label was available during annotation to allow individual annotators to explicitly mark clips for which they were uncertain about the presence or interpretation of sound events. Clips marked as ``Doubt'' were not immediately incorporated into the final ground truth and were systematically revisited every 10 active learning iterations (approximately every 10 weeks) for re-annotation, taking into account accumulated contextual information earned by the annotators.

\subsubsection{Active learning-based labeling procedure}

The labeling procedure of Soroll-IA is based on the active learning framework previously introduced in \cite{aes_berlin} and is tailored to the constraints of large-scale, real-world audio monitoring in outdoor industrial environments. Instead of relying on a single exhaustive annotation pass, labeling is conducted through iterative refinement cycles that combine automatic audio tagging with expert-driven validation, enabling efficient and consistent weak labeling under realistic operational conditions.

In each iteration, an audio tagging model trained on the current version of the dataset is used to infer clip-level predictions over unlabeled or partially labeled audio recordings. These predictions are subsequently analyzed to identify informative samples, including clips with low model confidence, conflicting class activations, or acoustically complex mixtures of events. Such samples are prioritized for manual inspection, following the uncertainty- and diversity-driven selection strategy described in \cite{aes_berlin}.

Domain experts then review the selected clips and assign weak, multilabel annotations, indicating the presence of sound events within each audio segment without specifying temporal boundaries. This process leverages both model feedback and expert knowledge of port operations and environmental acoustics, ensuring that labels remain semantically meaningful and operationally relevant. The validated annotations are incorporated into the ground truth, and the model is retrained, progressively improving its ability to capture rare events and reduce annotation noise.

This iterative active learning strategy was executed over 24 weekly iterations, resulting in a total of 9,600 annotated clips. Each iteration followed a fixed operational cycle: annotators performed labeling from Monday to Friday, processing 400 new clips per week distributed between the two annotator groups (200 per group). At the end of each week, the updated annotations were used to retrain the active learning model on Friday, and the updated model was then used to select the next batch of 400 clips, which was made available for annotation the following Monday. The process ensures that all clips are reviewed by domain-experienced annotators across groups, improving coverage and consistency of the labeling process.

This iterative active learning strategy enables Soroll-IA to achieve high annotation quality while controlling labeling effort, and naturally supports the generation of multiple ground-truth variants with different consensus criteria. By using the framework introduced in \cite{aes_berlin}, the labeling procedure ensures methodological continuity between the recording node, the data acquisition pipeline, and the final dataset release.

\subsection{Labeling Post Process}\label{subsec:postproces}

After completing all active learning labeling cycles, an additional post-processing stage was applied to the annotations to ensure consistency with the intended scope of the public release of Soroll-IA. This phase focused on refining the label set and removing content that did not correspond to industrial-port acoustic activity.

First, all audio clips that had been assigned any label related to music during the manual annotation process were systematically discarded. This initial filtering step removed clips containing background music or incidental musical content that could introduce semantic ambiguity and domain leakage into the dataset.

As a second validation pass, the remaining audio clips were re-evaluated using a CNN14 model from the PANNs family, pretrained on AudioSet. The model was used to estimate the presence of music-related content across the dataset. Any clip for which the predicted probability of the Music class exceeded 10\% was flagged for manual review and subsequently re-annotated by the labeling team. This assisted verification step allowed the identification and correction of residual music-related annotations that may have been overlooked during the initial labeling phase, while avoiding fully automated exclusion decisions.

This two-stage post-processing strategy—combining expert annotations with automated model-based filtering—ensures that Soroll-IA is strictly composed of industrial and port-related sound events, minimizing cross-domain contamination. As a result, the released dataset provides a cleaner and more coherent benchmark for research in audio tagging and weakly supervised sound event analysis in real-world industrial environments.

In addition, only sound event classes with at least 200 annotated audio clips were included in the final public release. This criterion was adopted to ensure a minimum level of statistical representativeness per class while preserving the inherent imbalance of real-world port soundscapes. Following the annotation process, classes containing fewer than 200 clips were found to be sparsely represented, making them less suitable for reliable statistical characterization and machine learning evaluation. Therefore, 200 clips was selected as a practical threshold balancing class diversity and data availability. As a result, the final dataset reflects both the long-term acoustic activity of the monitored port environment and the selection effects introduced by the active learning-based labeling strategy.

\subsection{Label taxonomy}\label{subsec:taxonomy}

A detailed description of the 26 sound classes is provided in Appendix~\ref{app:taxonomy}, Table~\ref{tab:soroll_classes}.

\section{Soroll-IA}\label{sec:sorollia}

Soroll-IA is a port-focused environmental audio dataset composed of 7,396 audio clips, each with a fixed duration of 10 seconds, resulting in a total of approximately 22 hours of recorded audio. All recordings were captured in real-world industrial port conditions using fixed outdoor sensing nodes. The dataset is annotated following a weak labeling paradigm at the clip level and is organized according to a taxonomy of 26 sound event classes, specifically designed to reflect typical acoustic events in port logistics and industrial operations.

Given that each audio clip was independently annotated by multiple expert labelers, Soroll-IA is released with two complementary ground-truth configurations. The first, referred to as Non–Cross-Validated (Non-CV) ground truth, aggregates all labels assigned by the annotators, such that the final annotation for a clip corresponds to the union of all sound events identified by any labeler. This setting preserves maximum label recall and reflects the inherent subjectivity and ambiguity of complex acoustic scenes. The second configuration, denoted as Cross-Validated (CV) ground truth, enforces an inter-annotator agreement criterion: only those sound event labels supported by at least two-thirds of the annotators are retained. Specifically, a consensus label was assigned such that a sound event was included in the ground truth when at least 2 out of 3 annotators agreed on its presence; for clips annotated by two annotators, full agreement was required. This consensus-based ground truth provides a more conservative annotation set, prioritizing label reliability and reducing potential annotation noise. 

% Given that each audio clip was independently annotated by multiple expert labelers, Soroll-IA is released with two complementary ground-truth configurations. The first, referred to as Non–Cross-Validated (Non-CV) ground truth, aggregates all labels assigned by the annotators, such that the final annotation for a clip corresponds to the union of all sound events identified by any labeler. This setting preserves maximum label recall and reflects the inherent subjectivity and ambiguity of complex acoustic scenes. The second configuration, denoted as Cross-Validated (CV) ground truth, enforces an inter-annotator agreement criterion: only those sound event labels supported by at least two-thirds of the annotators are retained. This consensus-based ground truth provides a more conservative annotation set, prioritizing label reliability and reducing potential annotation noise.

% \textcolor{blue}{A consensus label was assigned using a two-thirds agreement criterion. Specifically, a sound event was included in the ground truth when at least 2 out of 3 annotators agreed on its presence; for clips annotated by two annotators, full agreement was required.}}

All the dataset exploration, statistical analysis, and the 5-fold cross-validation configuration presented in this paper were obtained on the Non–Cross-Validated (Non-CV) ground truth. This choice enables a faithful representation of the full annotation space and avoids introducing early biases during dataset partitioning. Once the folds are defined under the Non-CV configuration, the Cross-Validated (CV) ground truth is derived as a refinement step, in which label consensus is reassessed within each predefined partition. Specifically, the CV configuration retains only those sound event labels that reach a two-thirds inter-annotator agreement, ensuring consistency across folds while improving label reliability. This two-stage strategy decouples data partitioning from consensus enforcement, preserving both statistical balance and annotation rigor.

\begin{figure}
    \centering
    \includegraphics[width=1\linewidth]{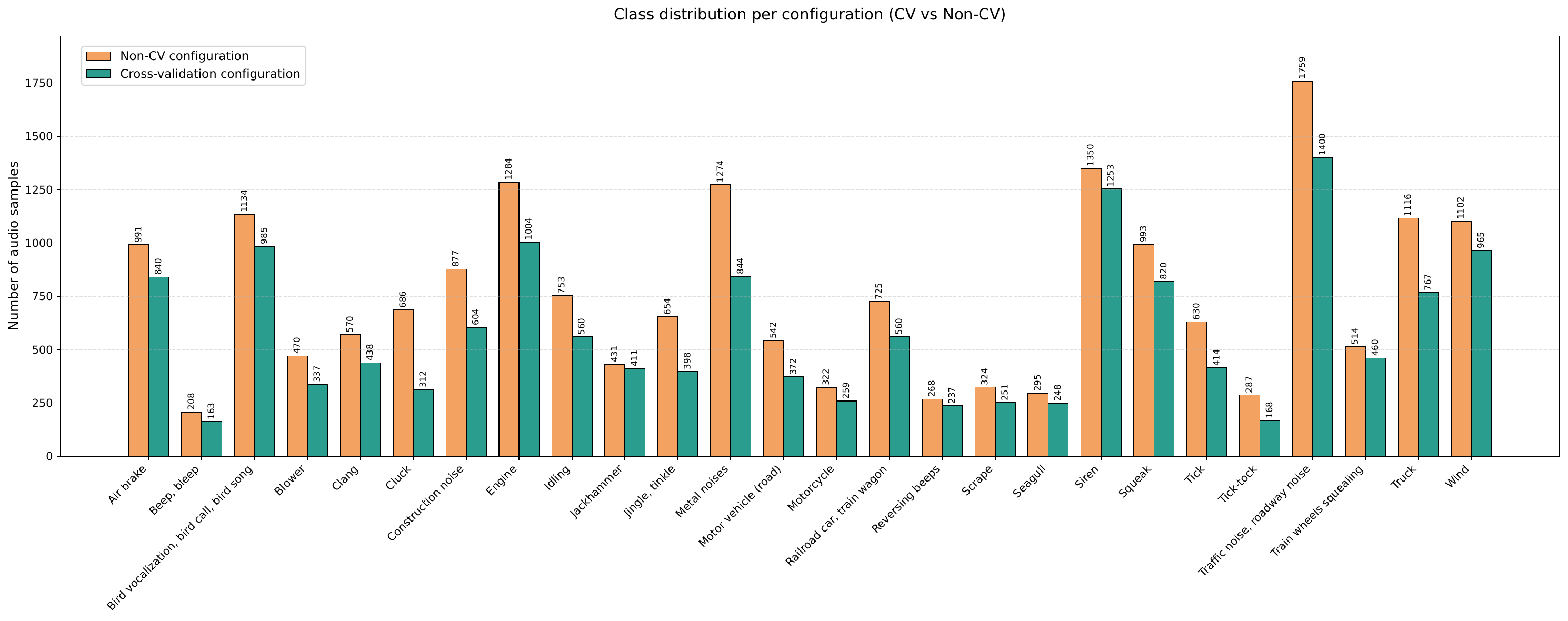}
    \caption{Event class Soroll-IA distribution in both configurations}
    \label{fig:final_hist}
\end{figure}

Fig.~\ref{fig:final_hist} illustrates the class-wise distribution of audio clips in Soroll-IA under the two released ground-truth configurations: Non-CV and CV. In both cases, the histogram reports the number of audio clips in which each class is present, reflecting the weak, multilabel nature of the annotations.

A clear class imbalance is observed across the taxonomy in both configurations, which is consistent with real-world industrial port soundscapes. Frequent and persistent sound sources—such as Traffic noise (roadway), Engine, Siren, Wind, Bird vocalization, and Truck—exhibit the highest number of occurrences. These sounds are characteristic of continuous or recurrent port operations and therefore accumulate a large number of positive clips over long-term recording. In contrast, short-duration or sporadic events—such as Beep, Tick-tock, Motorcycle, or Scrape—appear significantly less often, highlighting the natural skew induced by operational dynamics rather than artificial dataset curation.

Comparing the Non-CV and CV configurations, the CV counts are systematically lower, as expected, since only labels with a two-thirds annotator consensus are retained. However, the relative ordering of classes remains largely preserved across configurations, indicating that the cross-validation process refines label reliability without altering the overall acoustic profile of the dataset.

Importantly, some classes show high agreement between annotators, reflected by relatively small differences between Non-CV and CV counts. These include acoustically salient and easily identifiable events such as Jackhammer, Siren, Reversing Beeps, Train wheels squealing, Seagull and Wind. Their strong acoustic signatures and longer temporal footprints facilitate consistent labeling across annotators.

Conversely, other classes exhibit a larger relative drop from Non-CV to CV, suggesting lower inter-annotator consensus. These are typically short, transient, or acoustically ambiguous events—such as Cluck, Jingle tinkle, or Tick—which may be partially masked by background noise or overlap with other sound sources. This behavior underscores the intrinsic challenges of weak labeling in complex outdoor industrial environments and motivates the release of both ground-truth configurations.

Overall, Fig~\ref{fig:final_hist} highlights that Soroll-IA faithfully captures both the acoustic diversity and the annotation uncertainty inherent to real-world port environments, offering researchers the flexibility to select between a more inclusive label set (Non-CV) and a more conservative, high-consensus annotation scheme (CV), depending on the target application.

Importantly, the observed class distribution is also influenced by the use of an active-learning–driven labeling strategy that prioritizes informative and rare samples but does not enforce uniform class frequencies. Consequently, common events such as traffic-related noise, engine sounds, or crane sirens dominate the dataset, while less frequent events—although still represented above a minimum occurrence threshold—remain underrepresented. This characteristic makes Soroll-IA particularly suitable for evaluating audio tagging methods under realistic data imbalance conditions, closely mirroring the challenges faced by deployed industrial audio monitoring systems. 

\begin{figure}
    \centering
    \includegraphics[width=1\linewidth]{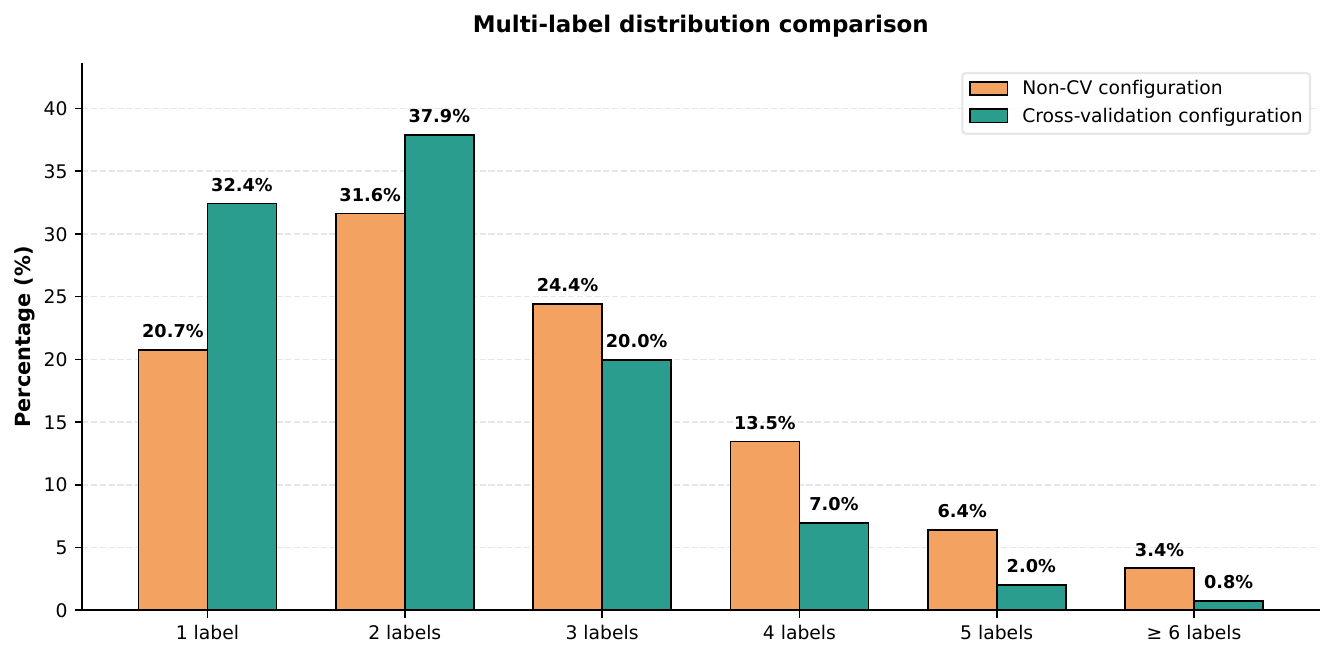}
    \caption{Comparison of the multi-label distribution between the non-cross-validation (Non-CV) and cross-validation configurations. The bars represent the percentage of samples corresponding to each number of labels. Samples with six or more labels are grouped into the ``$\geq$ 6 labels" category. }
    \label{fig:final_multi_label}
\end{figure}
%Total sample sizes for each configuration ($N$) are indicated in the legend.

To provide a rigorous quantitative assessment of the dataset's polyphonic nature and validate its multi-label complexity, we analyzed the distribution of concurrent acoustic events per sample. Fig.~\ref{fig:final_multi_label} illustrates the percentage of audio excerpts categorized by the number of active labels, comparing the Non-CV and Cross-Validation configurations. The empirical distribution demonstrates that single-label samples constitute only a minor fraction of the data (20.7\% for Non-CV and 32.4\% for Cross-Validation), meaning that the vast majority of the instances exhibit a high degree of acoustic overlap (79.3\% and 67.6\%, respectively). Specifically, instances with 2 or 3 concurrent labels represent the dominant profiles in both setups. Furthermore, a non-negligible portion of the data contains highly dense environments featuring 4, 5, or even 6 or more classes ($\geq 6$ labels). When comparing both setups, a slight increase in the proportions of 1 and 2-label samples is observed in the Cross-Validation configuration, whereas higher-order polyphonic instances ($\geq 3$ labels) decrease accordingly. This shift is due to the data curation process required for the Cross-validation setup, where audio samples lacking annotator consensus or containing highly ambiguous overlapping sources were discarded. This filtering naturally pruned some of the most complex, multi-labeled instances, leading to a higher concentration of cleaner 1 and 2-label samples. Nevertheless, the Cross-Validation configuration preserves the overall polyphonic signature of the Non-CV split, ensuring that the intrinsic difficulty of the multi-label audio tagging task remains consistent across all cross-validated folds. This high density of mutilabel samples directly increases task complexity, as different sources obscure individual acoustic signatures, thereby justifying the dataset's challenge for real-world deployment benchmarks.

\begin{figure}
    \centering
    \includegraphics[width=1\linewidth]{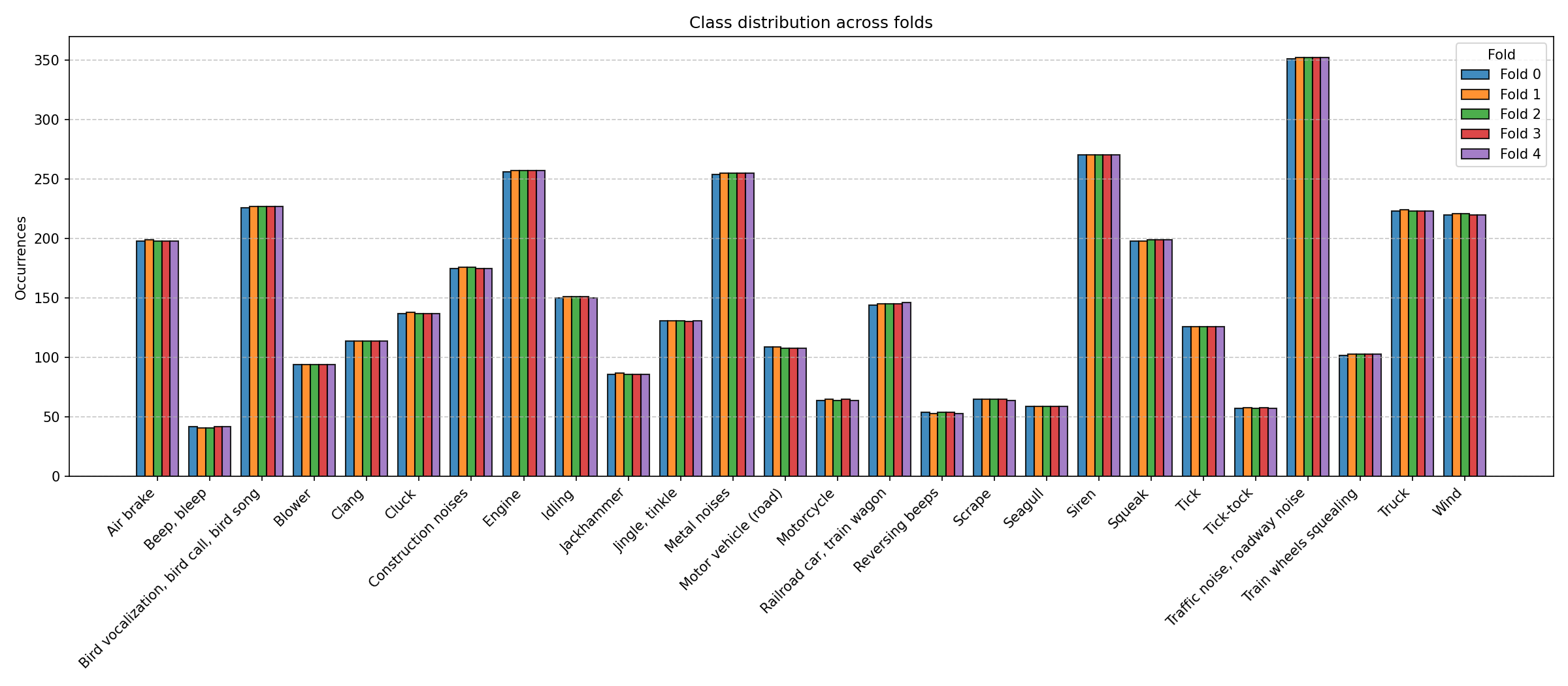}
    \caption{Distribution of classes per folds in Non-CV configuration}
    \label{fig:kfold}
\end{figure}

Fig.~\ref{fig:kfold} illustrates the distribution of sound event classes across the five folds of the Soroll-IA dataset with Non-CV configuration. As can be observed, the class frequencies are consistently preserved across all folds, ensuring that each partition reflects the overall class distribution of the dataset. 

This balanced allocation is particularly important given the inherent class imbalance present in real-world port and industrial soundscapes. By maintaining comparable class occurrences in each fold, the proposed k-fold configuration minimizes evaluation bias and prevents performance variations caused by uneven class representation rather than model behavior. Consequently, this setup enables fair and reproducible comparisons across different models, training strategies, and experimental conditions, while remaining faithful to the acoustic reality captured during long-term monitoring at the port.

Soroll-IA constitutes a real-world, outdoor industrial audio dataset specifically designed to address the lack of publicly available resources for sound event tagging in a port environment. Captured over long-term deployments using fixed sensing nodes and curated through an expert-driven active learning pipeline, the dataset reflects the acoustic complexity, class imbalance, and event overlap inherent to operational ports. Its focus on weak multilabel annotations, industrially relevant sound classes, and a k-fold evaluation protocol makes Soroll-IA a robust and realistic benchmark for developing and evaluating computer audition methods, particularly those targeting low-latency, edge-based, and industrial monitoring applications.

\section{Baseline/Benchmark}\label{sec:baseline}

% Two benchmark models are considered in this study: one based on the state-of-the-art CNN14 architecture from the PANNs family, and a second model built upon MobileNetV2. Both models were trained from scratch using a sampling rate of 32 kHz, strictly following the training procedures defined in the official PANNs repository\footnote{\url{https://github.com/qiuqiangkong/audioset_tagging_cnn}}. Training was carried out for 100 epochs, and all experiments were conducted on a Tesla V100S GPU with 32 GB of memory.

Two benchmark models were considered in this study: the state-of-the-art CNN14 architecture from the PANNs family and a lightweight MobileNetV2-based model. Both architectures were evaluated under two training strategies: (i) training from scratch and (ii) fine-tuning from the official AudioSet-pretrained checkpoints provided in the PANNs repository\footnote{\url{https://github.com/qiuqiangkong/audioset_tagging_cnn}}. To ensure a fair comparison, all experiments followed a unified training protocol. Audio signals were resampled to 32 kHz and converted into log-mel spectrograms using a 1024-point window, a hop size of 320 samples, and 64 mel frequency bins covering the frequency range from 50 Hz to 14 kHz. Models were trained using a clip-level binary cross-entropy loss without class balancing, and mixup data augmentation was applied to improve generalization. Training was performed with the Adam optimizer, a learning rate of 0.001, and a batch size of 32 for a total of 100 epochs. All experiments were conducted on a Tesla V100S GPU with 32 GB of memory. Baseline code can be found in this repository\footnote{\url{https://github.com/anp-iti/sorollia_baseline}}

% Añadir?
% Both benchmark models, CNN14 and MobileNetV2, were trained from scratch following a unified training protocol to ensure a fair comparison. Audio signals were resampled to 32 kHz and converted into log-mel spectrograms using a 1024-point window, a hop size of 320 samples, and 64 mel frequency bins, covering the frequency range from 50 Hz to 14 kHz. Training was performed using a clip-level binary cross-entropy loss, without class balancing, and mixup data augmentation was applied to improve generalization. Models were trained with a batch size of 32 and a learning rate of 0.001 for a total of 100 epochs, using the Adam optimizer. All experiments were conducted on a Tesla V100S 32GB GPU.

%This approach ensures a fair comparison between the models and aligns with best practices for audio event classification tasks.

%The motivation behind this benchmarking approach is to provide a clear starting point for future research by using a well-established and widely adopted model in the computational audio community, namely CNN14 from the PANNs family \cite{khandelwal2023leveraging, koh2022automated, liu2022leveraging, xiao2023graph}. This model serves as a robust reference due to its strong performance and extensive validation in audio tagging tasks. .

The motivation behind this benchmarking approach is to provide a clear starting point for future research by using a well-established and widely adopted model in the computational audio community, namely CNN14 from the PANNs family \cite{khandelwal2023leveraging, koh2022automated, liu2022leveraging, xiao2023graph}. This model serves as a robust reference due to its strong performance and extensive validation in audio tagging tasks. To comprehensively assess its capabilities on the proposed dataset, two distinct training approaches were evaluated: first, training the architecture entirely from scratch to establish a baseline tailored solely to our data (see Section~\ref{subsec:cnn14}); and second, leveraging transfer learning through a fine-tuning strategy (see Section~\ref{subsec:cnn14_finetune}) using weights pre-trained on AudioSet \cite{gemmeke2017audio}. In parallel, we propose a second benchmark based on MobileNetV2 (trained from scratch), chosen for its lightweight architecture and low parameter count. This enables the exploration of real-time audio tagging solutions tailored for embedded systems, where computational resources and power consumption are critical constraints. Together, these benchmarks offer a comprehensive baseline covering both high-performance and resource-efficient approaches.

To facilitate robust and unbiased evaluation, the Soroll-IA dataset is released using a 5-fold cross-validation (k-fold) configuration. This design choice aims to mitigate potential biases that may arise from relying on a single, fixed train/test split, particularly in long-term real-world recordings where temporal correlations and class imbalance are inherent.

By rotating the training and evaluation subsets across five folds, each audio clip is used for both training and validation in different iterations, enabling a more reliable estimation of model generalization performance. This setup is especially relevant for port environments, where acoustic conditions evolve over time and sound event distributions are highly non-uniform. The k-fold configuration ensures that performance comparisons across different models and experimental settings are fair, reproducible, and less sensitive to the specific characteristics of a single partition.

\subsection{Scratch-Based CNN14 State of the art Benchmarking}\label{subsec:cnn14}

\begin{table}[tbp]
\centering
\caption{Comparison of performance metrics between CV and Non-CV benchmarks with CNN14 model}
\begin{tabular}{lcccc}
\toprule
\textbf{Metric} & \textbf{Benchmark} & \textbf{Mean} & \textbf{Std} & \textbf{95\% CI} \\
\midrule
\multirow{2}{*}{mAP} & Non-CV & 0.6700 & 0.0075 & [0.6634, 0.6766] \\
                     & CV     & 0.6358 & 0.0205 & [0.6179, 0.6537] \\
\midrule
\multirow{2}{*}{macro F1} & Non-CV & 0.5770 & 0.0205 & [0.5590, 0.5950] \\
                          & CV     & 0.5338 & 0.0259 & [0.5111, 0.5566] \\
\bottomrule
\end{tabular}
\label{tab:benchmark_comparison}
\end{table}

% Table~\ref{tab:benchmark_comparison} presents a comparison of the mean Average Precision (mAP) and macro F1-score between the Non-Cross-Validation (Non-CV) and Cross-Validation (CV) setups. The Non-CV results show better performance with a higher mAP of 0.670 and macro F1 of 0.577, compared to 0.636 and 0.534 for the CV approach, respectively. These metrics reflect the model’s effectiveness in detecting and classifying audio events under different evaluation strategies.

Table~\ref{tab:benchmark_comparison} presents a comparison of the mean Average Precision (mAP) and macro F1-score between the Non-Cross-Validation (Non-CV) and Cross-Validation (CV) setups. The Non-CV configuration yields higher metric values, with an mAP of 0.670 and a macro F1-score of 0.577, compared to 0.636 and 0.534 for the CV setup, respectively. This difference is expected due to the more permissive labeling strategy in the Non-CV setting, which aggregates annotations across annotators and increases label coverage. In contrast, the CV configuration provides a more conservative and robust estimate of generalization performance under stricter evaluation conditions. These results should therefore be interpreted in terms of evaluation protocol differences rather than as a direct indication of superior model behavior.

Standard deviations and 95\% confidence intervals (CI) are also reported to provide an estimate of variability and statistical uncertainty around the mean values. The inclusion of the 95\% CI helps to understand the precision of the results and to assess whether performance differences are likely to be meaningful.

\begin{figure}[tbp]
    \centering
    \includegraphics[width=1\linewidth]{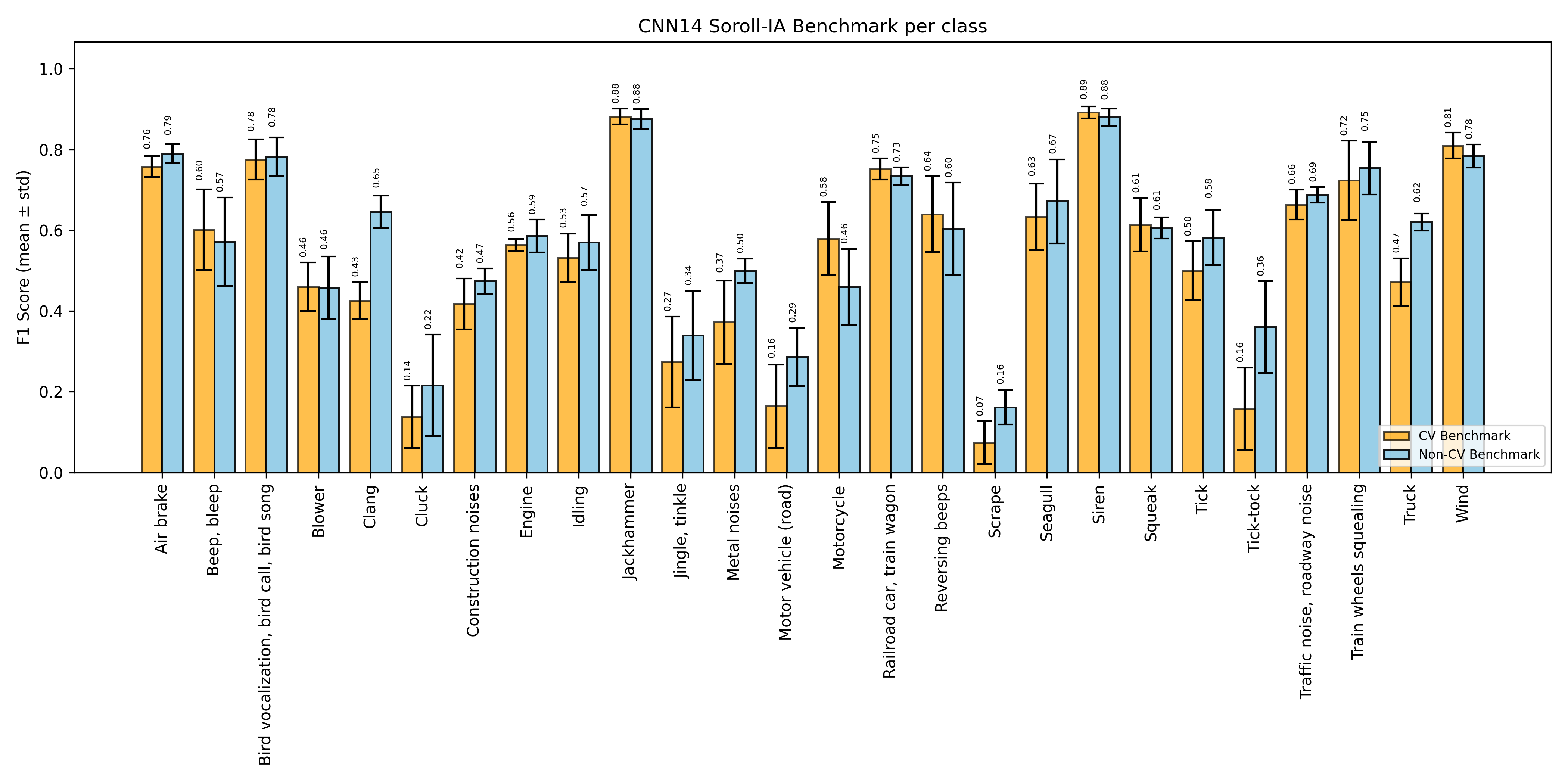}
    \caption{CNN14 Benchmark for both configurations}
    \label{fig:cnn14class}
\end{figure}

% Overall, the Non-CV setup appears to yield more stable and higher performance metrics, while the CV setup offers a more conservative but realistic estimate of model generalization across data splits.

Overall, the Non-CV setup yields higher metric values, while the CV setup provides a more conservative and robust estimate of model generalization across data splits. The observed differences are primarily driven by the more permissive labeling aggregation in the Non-CV configuration compared to the stricter evaluation protocol used in CV.

Fig.~\ref{fig:cnn14class} reports the per-class F1 scores (mean ± standard deviation across folds) obtained by the CNN14 model from the PANNs family when evaluated on Soroll-IA under both the Non-CV and cross-validated (CV) ground-truth configurations. Overall, the results reveal a strong dependency between class performance, acoustic characteristics, and annotation consensus.

Across most classes, Non-CV benchmarks achieve equal or higher F1 scores than their CV counterparts. This behavior is expected, as the Non-CV configuration aggregates all annotator-provided labels, resulting in a more inclusive ground truth that tolerates partial agreement. In contrast, the CV configuration enforces a stricter two-thirds consensus, increasing label precision at the cost of reduced positive samples, which can negatively affect recall—particularly for rare or ambiguous events.

Classes characterized by distinctive and persistent acoustic patterns consistently achieve high F1 scores in both configurations. Notable examples include Jackhammer, Siren, Traffic noise (roadway), Wind, Engine, and Train wheels squealing, with F1 values often exceeding 0.75 and relatively low variance. These sounds exhibit strong spectral signatures and longer temporal presence within clips, facilitating robust detection even under weak labeling.

Conversely, short-duration or transient events—such as Cluck, Tick, Tick-tock and Scrapes—present significantly lower F1 scores, even under the Non-CV configuration, with further degradation observed under CV settings. These classes suffer from both acoustic sparsity and higher inter-annotator disagreement, as previously observed in the class distribution analysis (see Fig.~\ref{fig:final_hist}). Their performance is consistently unstable across folds, as reflected by larger standard deviations, indicating intrinsic difficulty independent of the validation strategy. Future work may investigate strategies tailored to imbalanced and low-resource acoustic classes, including class-weighted loss functions, focal loss, and data augmentation techniques adapted to transient sound events, which may improve robustness when working with the presented dataset.

%Conversely, short-duration or transient events—such as Cluck, Tick, Tick-tock and Scrapes—present significantly lower F1 scores, particularly under the CV configuration. These classes suffer from both acoustic sparsity and higher inter-annotator disagreement, as previously observed in the class distribution analysis. The larger standard deviations for these classes further indicate unstable model performance across folds, highlighting their inherent difficulty. 
%\textcolor{blue}{Future work may investigate strategies specifically designed for imbalanced and low-resource classes, including class-weighted loss functions, focal loss, and targeted data augmentation techniques, which may help improve robustness for these underrepresented sound events.}

Intermediate performance is observed for classes like Blower, Clang, Construction noises, Metal noises, and Motorcycle, where F1 scores typically range between 0.4 and 0.65. For these categories, the performance gap between Non-CV and CV configurations is moderate, suggesting partial annotator consensus and a reasonable balance between label noise and model generalization. Interestingly, some classes—such as Seagull and Bird vocalization, bird call, bird song—maintain competitive F1 scores despite not being strictly industrial sounds.

Overall, Fig.~\ref{fig:cnn14class} demonstrates that annotation strictness directly impacts benchmark outcomes, particularly for rare and ambiguous events. The availability of both CV and Non-CV configurations therefore, provides a valuable experimental axis: the Non-CV setup reflects performance under inclusive, realistic labeling conditions, while the CV configuration offers a conservative benchmark emphasizing high-confidence annotations. Together, they enable a nuanced evaluation of audio tagging systems in complex industrial port environments.

\begin{table}[tbp]
\centering
\caption{Comparison of performance metrics between CV and Non-CV benchmarks with AudioSet pre-trained CNN14 model under different fine-tuning strategies.}
\begin{tabular}{lllccc}
\toprule
\textbf{Metric} & \textbf{Benchmark} & \textbf{Strategy} & \textbf{Mean} & \textbf{Std} & \textbf{95\% CI} \\
\midrule

\multirow{4}{*}{mAP}
& \multirow{2}{*}{Non-CV} & Classifier only &  0.5604 & 0.0068 & [0.5520, 0.5688] \\
&                         & Embedding + Classifier &  0.6366 & 0.0072 & [0.6277, 0.6455] \\
\cmidrule(lr){2-6}
& \multirow{2}{*}{CV}     & Classifier only & 0.5396 & 0.0056 & [0.5327, 0.5465] \\
&                         & Embedding + Classifier & 0.6136 & 0.0053 & [0.6070, 0.6202] \\
\midrule

\multirow{4}{*}{Macro F1}
& \multirow{2}{*}{Non-CV} & Classifier only & 0.3200 & 0.0122  & [0.3048, 0.3352] \\
&                         & Embedding + Classifier & 0.4980 & 0.0110 &  [0.4844, 0.5116] \\
\cmidrule(lr){2-6}
& \multirow{2}{*}{CV}     & Classifier only & 0.2880 & 0.0045 & [0.2824, 0.2936] \\
&                         & Embedding + Classifier & 0.4769 & 0.0111 &  [0.4649, 0.4871] \\
\bottomrule
\end{tabular}
\label{tab:benchmark_comparison_finetune}
\end{table}

\subsection{AudioSet Pre-trained CNN14 Fine-Tuning Benchmarking}\label{subsec:cnn14_finetune}

This section introduces a fine-tuning benchmarking setup based on the AudioSet pre-trained CNN14 model from the PANNs family. The objective is to evaluate the impact of different transfer learning strategies on the downstream audio tagging performance of Soroll-IA under both CV and Non-CV evaluation protocols. Two fine-tuning configurations are considered. The first strategy updates only the final classification (tagging) layer, while keeping the rest of the network frozen. The second strategy extends the trainable parameters to include both the penultimate embedding layer and the final classification layer, allowing partial adaptation of the high-level representations to the target domain.

Table~\ref{tab:benchmark_comparison_finetune} presents the performance comparison between the CV and Non-CV benchmarks under the two evaluated fine-tuning strategies. The experimental results reveal two distinct and critical trends. First, unlocking the penultimate embedding layer alongside the classification layer (Embedding + Classifier) yields a substantial and statistically significant performance boost across all metrics and evaluation protocols compared to training the classification layer alone (Classifier only). Specifically, for the Non-CV benchmark, the mean mAP increases from 0.5604 to 0.6366, while the Macro F1 score experiences a remarkable surge from 0.3200 to 0.4980. A homologous behavioral pattern is observed in the CV setup, where mAP improves by approximately 7.4 percentage points and Macro F1 increases from 0.2880 to 0.4769. This consistent improvement demonstrates that while AudioSet features provide a robust general baseline, its default pre-trained weights are insufficient to fully capture the specific, high-density acoustic profiles of the Soroll-IA target domain due to a severe domain mismatch. In fact, this discrepancy is further evidenced by the superior performance achieved when training the architecture entirely from scratch (as detailed in Section~\ref{subsec:cnn14}), which underscores that industrial and port soundscapes possess highly localized signatures that general-purpose audio databases fail to model adequately. This necessity for domain-specific feature adaptation or custom training strategies aligns with our previous observations regarding acoustic density and industrial noise characterization \cite{Garcia-Ballesteros2026confidently}.

\begin{figure*}[tbp]
\centering
\includegraphics[width=\textwidth]{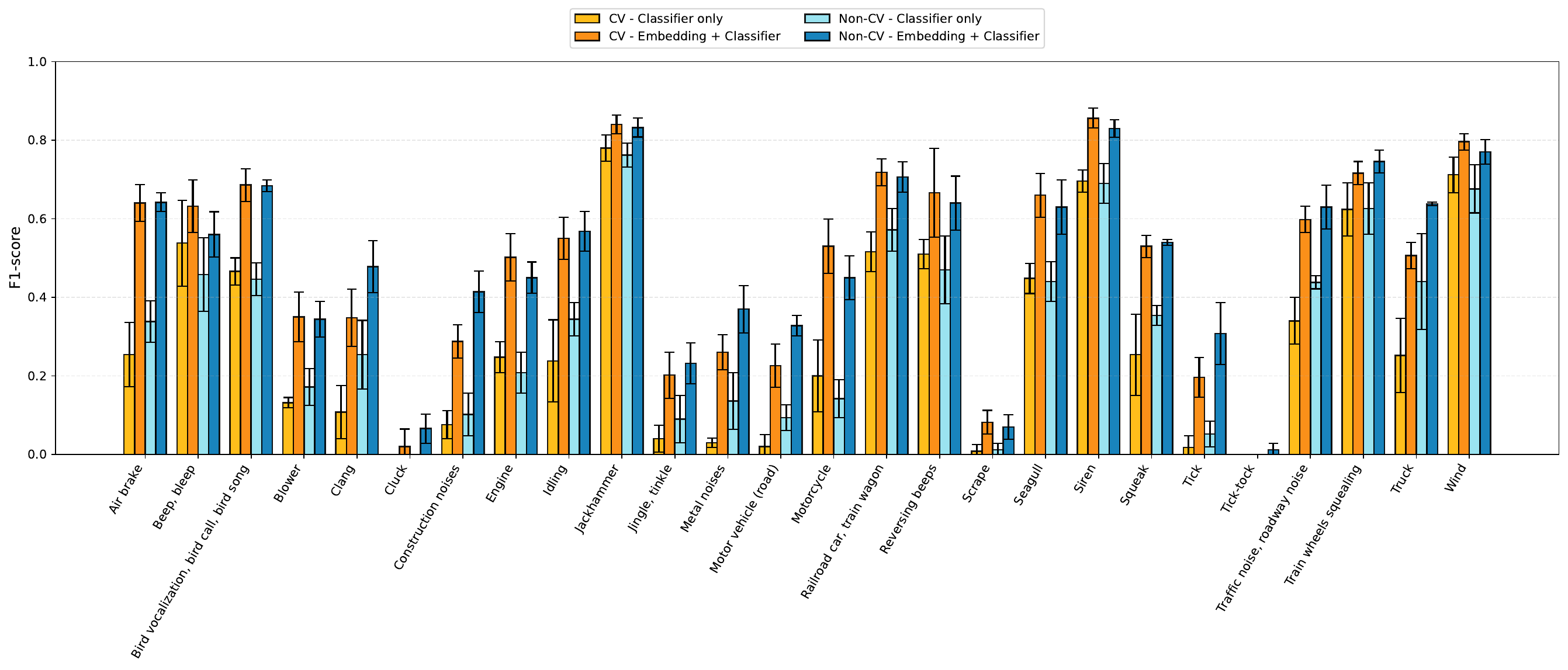}
\caption{Per-class F1-score comparison across the evaluated fine-tuning strategies (Classifier only vs. Embedding + Classifier) under both Non-CV and CV experimental benchmarks. Error bars indicate the 95\% confidence intervals.}
\label{fig:per_class_finetune}
\end{figure*}

To uncover the granular impact of the transfer learning strategies, a per-class F1-score analysis was conducted (Fig.~\ref{fig:per_class_finetune}). The results reveal that adapting the embedding space (Embedding + Classifier) is highly critical for specialized industrial sounds like Air brake, Idling, and Truck, where F1-scores substantially increase or even double compared to the frozen baseline (Classifier only). Conversely, distinctive acoustic categories with robust time-frequency profiles, such as Siren, Jackhammer, and Wind, maintain consistently high performance ($\geq 0.70$) across all configurations, indicating that their intrinsic signatures transfer effectively from the source domain. Finally, rare or highly transient classes (e.g., Cluck and Tick-tock) remain a persistent challenge with F1-scores near zero, confirming that background polyphony and class imbalance represent the core bottlenecks of this benchmark.

% This specific trend provides definitive empirical evidence of a severe domain mismatch between AudioSet and Soroll-IA, demonstrating that general-purpose pre-trained weights fail to natively capture the localized, high-density acoustic signatures of outdoor port environments without partial fine-tuning.

\subsection{MobileNetV2 low parameter baseline}\label{subsec:mobilenet}

Table \ref{tab:benchmark_comparison_mobilenetv2} compares the performance metrics of the MobileNetV2 model under the two released evaluation setups. MobileNetV2 serves as a lightweight baseline model designed for IoT applications, where computational resources are limited.

The mean Average Precision (mAP) is higher in the Non-CV benchmark (0.6522) compared to the CV setup (0.6252), suggesting that evaluation without cross-validation may overestimate model performance due to less variability between training and testing splits.

Similarly, the macro F1 score shows a higher mean in the Non-CV benchmark (0.5754) than in CV (0.5439). The Non-CV results also exhibit lower standard deviations and narrower 95\% confidence intervals, indicating less variability in the performance estimates.

% Overall, while the Non-CV setup yields seemingly better results, the CV evaluation provides a more robust and realistic assessment of the model’s generalization capabilities, which is critical for deploying efficient models like MobileNetV2 in resource-constrained IoT environments.

\begin{table}[tbp]
\centering
\caption{Comparison of performance metrics between CV and Non-CV benchmarks with MobileNetV2 model}
\begin{tabular}{lcccc}
\toprule
\textbf{Metric} & \textbf{Benchmark} & \textbf{Mean} & \textbf{Std} & \textbf{95\% CI} \\
\midrule
\multirow{2}{*}{mAP} & Non-CV & 0.6522 & 0.0068 & [0.6463, 0.6581] \\
                     & CV     & 0.6252 & 0.0067 & [0.6193, 0.6311] \\
\midrule
\multirow{2}{*}{macro F1} & Non-CV & 0.5754 & 0.0134 & [0.5636, 0.5872] \\
                          & CV     & 0.5439 & 0.0174 & [0.5286, 0.5592] \\
\bottomrule
\end{tabular}
\label{tab:benchmark_comparison_mobilenetv2}
\end{table}

% \textcolor{blue}{Overall, the Non-CV setup yields higher metric values due to its more permissive labeling definition, which aggregates annotations across annotators and therefore increases label coverage. In contrast, the CV evaluation provides a more conservative and robust estimate of generalization performance, as it evaluates the model under stricter agreement conditions. This distinction is important for interpreting the results, as the observed differences primarily reflect variations in annotation aggregation and evaluation strictness rather than differences in intrinsic model behavior. The CV setting is therefore the most reliable indicator of expected performance in deployment scenarios, particularly for resource-constrained IoT environments where generalization is critical.}

Consistent with the trends observed in the CNN14 benchmarks (see Sections~\ref{subsec:cnn14}~and~\ref{subsec:cnn14_finetune}), the MobileNetV2 results confirm a systematic performance drop when transitioning from the Non-CV to the stricter CV evaluation protocol. However, this gap is particularly informative for this lightweight architecture, as it demonstrates that MobileNetV2 maintains stable generalization capabilities despite its reduced parameter footprint. While the Non-CV setup benefits from aggregated label coverage, the CV framework provides the most realistic and rigorous performance baseline for edge deployment, proving that the model can reliably handle complex, multi-label industrial environments under constrained IoT hardware resources.

\begin{figure}[tbp]
    \centering
    \includegraphics[width=1\linewidth]{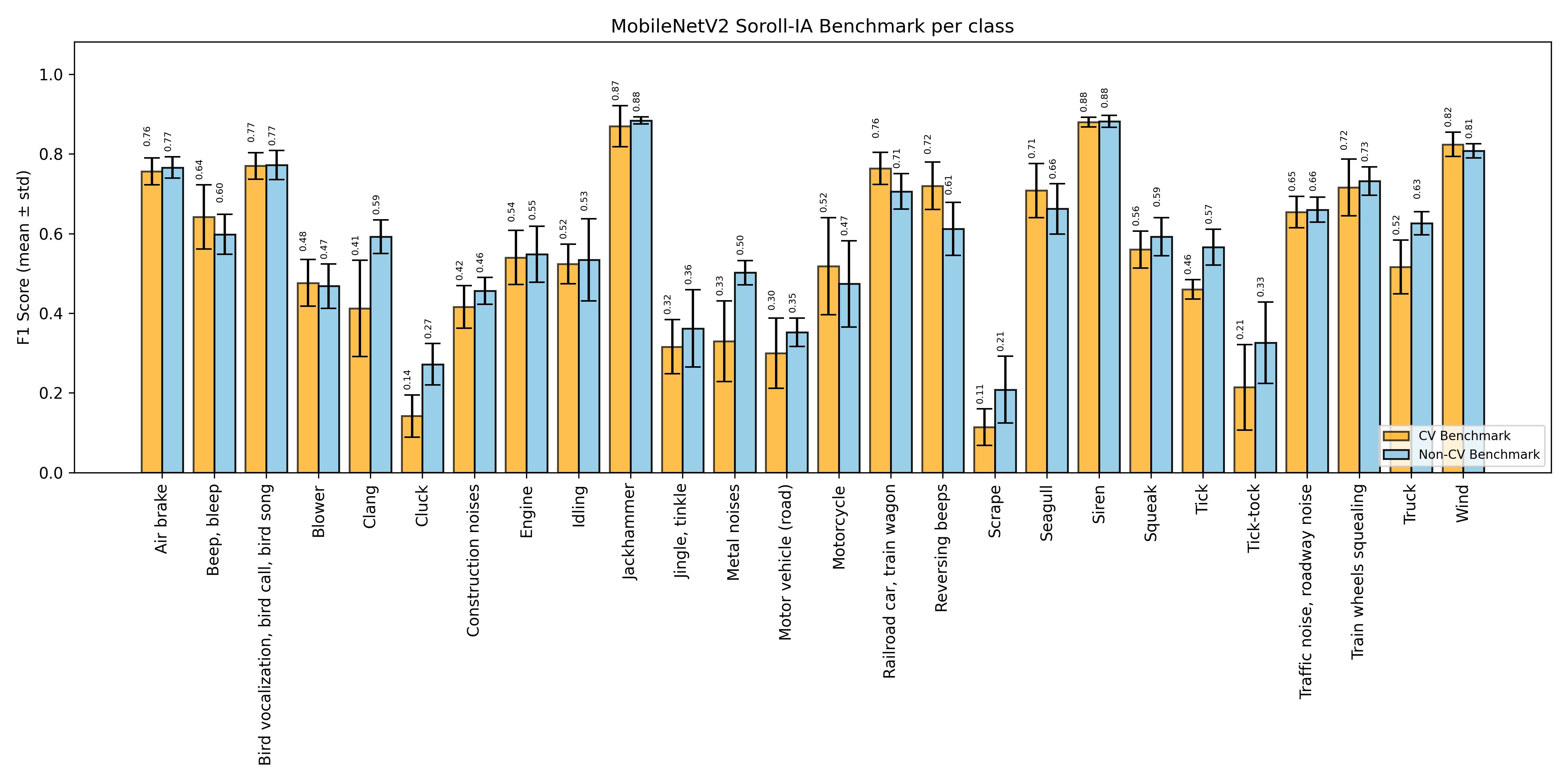}
    \caption{MobileNetV2 Benchmark for both configurations}
    \label{fig:mobilenetv2class}
\end{figure}

Fig.~\ref{fig:mobilenetv2class} presents the per-class macro F1-score performance of the MobileNetV2 benchmark on the Soroll-IA dataset, comparing the Cross-Validation (CV) and Non-Cross-Validation (Non-CV) evaluation setups. Overall, the Non-CV benchmark tends to achieve slightly higher F1-scores across most classes, which is consistent with the global metrics reported previously. High-performance classes such as Jackhammer, Siren, Wind, and Air brake exhibit strong and stable performance under both evaluation schemes, suggesting that these sound events are well represented and easier to discriminate. In contrast, classes such as Cluck and Tick-tock show lower F1-scores and higher variability, highlighting the increased difficulty of detecting infrequent or acoustically ambiguous events.

The CV setup generally results in more conservative performance estimates, with lower mean F1-scores and, in some cases, larger standard deviations, reflecting the increased robustness and variability introduced by data resampling. This behavior is particularly evident in mid-performing classes such as Motorcycle, Construction noises, and Scrape. These results emphasize the importance of cross-validation for obtaining reliable performance estimates, especially when evaluating lightweight models intended for deployment in real-world IoT and embedded audio tagging scenarios.

\FloatBarrier
\section{Conclusion}\label{sec:conclusion}

% This paper has presented Soroll-IA, the first publicly available environmental audio dataset specifically designed around the real-world acoustic conditions of an industrial port. Unlike existing urban or industrial datasets, Soroll-IA captures the operational, logistical, and environmental complexity of a port infrastructure, a domain that has remained largely unexplored in the computer audition literature despite its clear relevance for safety, logistics optimization, and environmental monitoring.

This paper has presented Soroll-IA, a publicly available environmental audio dataset focused on industrial port acoustics. The dataset is collected from a real-world industrial port in Valencia (Spain) and reflects the acoustic conditions of a Mediterranean port environment recorded over a one-year deployment using fixed sensing nodes. While the recordings are necessarily specific to this geographic location and time period, they capture a range of sound events and operational conditions that are commonly encountered in industrial ports, such as crane operations, train movements, and road traffic. The dataset is therefore intended as a first step towards a systematic study of port acoustics in computer audition, a domain that remains underrepresented in the literature despite its relevance for safety, logistics optimization, and environmental monitoring.

The dataset is the result of one full year of continuous recording using fixed outdoor sensing nodes deployed in two strategically distinct port locations. Each node reflects a different acoustic reality within the port: one positioned near the main entrance, influenced by road traffic, construction activities, and idling vehicles; and another located close to railway tracks and port cranes, dominated by events such as train wheel squealing and crane sirens. This multi-node setup enables Soroll-IA to capture a diverse range of sound event dynamics and operational contexts within the same industrial environment.

To address the challenges of large-scale annotation in such complex soundscapes, Soroll-IA was constructed through an iterative active-learning framework, allowing the dataset to evolve in a data-driven manner while progressively refining label quality. The released dataset includes two ground-truth configurations: a Non-CV version that aggregates all annotator labels to preserve maximum event coverage, and a Cross-Validated (CV) version that enforces a two-thirds inter-annotator consensus to enhance label reliability. This dual configuration provides flexibility for different research objectives, ranging from exploratory analysis to robust model evaluation.

% In addition to the dataset itself, two benchmark baselines are provided to establish reference performance levels under contrasting computational paradigms. The first benchmark leverages CNN14 from the PANNs family, representing a widely adopted high-capacity architecture in audio tagging research. The second benchmark is based on MobileNetV2, targeting low-parameter, real-time inference scenarios suitable for edge and embedded deployments—an essential requirement for practical port monitoring systems.

In addition to the dataset itself, two benchmark baselines are provided to establish reference performance levels under contrasting computational paradigms. The first benchmark leverages CNN14 from the PANNs family, representing a widely adopted high-capacity architecture in audio tagging research, evaluated both by training completely from scratch and through two transfer learning fine-tuning strategies (Classifier-only vs. Embedding + Classifier) using AudioSet weights to address domain mismatch. The second benchmark is based on MobileNetV2, targeting low-parameter, real-time inference scenarios suitable for edge and embedded deployments—an essential requirement for practical port monitoring systems.

Finally, Soroll-IA reflects the heterogeneous nature of real industrial-port soundscapes, encompassing both long-duration, quasi-stationary events such as trains and roadway traffic, and short, impulsive sounds such as Scrape or Tick. This variability poses significant challenges for audio tagging models and makes Soroll-IA a realistic and demanding testbed for future research.

By releasing Soroll-IA, along with its annotations, folds, and benchmarks, we aim to foster further advances in weakly supervised audio tagging, edge-oriented sound analysis, and real-time acoustic monitoring in industrial environments. We believe this dataset will serve as a valuable foundation for the development and evaluation of next-generation computer audition systems tailored to complex, safety-critical, and operationally relevant domains such as industrial ports.

\section*{Declarations}\label{sec:declarations}

\subsection*{Funding}

The participation of all the researchers in this work was funded by the Valencian Institute for Business Competitiveness (IVACE) and the FEDER funds by means of projects LIASound (IMDEEA/2024/110), SOROLL-IA (IMDEEA/2022/68) and SOROLL-IA2 (IMDEEA/2023/91). This work has been partially supported by the Spanish State Research Agency and the European Union through the STARRING project, grant \mbox{PID2022-137048OA-C44}, funded by \\ \mbox{MICIU/AEI/10.13039/501100011033} and by ERDF, EU.

% \subsection*{Conflict of interest/Competing interest}

% The authors declare that they have no competing interests

\subsection*{Availability of data and materials}

Data supporting this study are included within the article and/or supporting materials. The license under which the data are provided for this release is Attribution-NonCommercial 4.0 International.

\subsection*{Acknowledgments}\label{sec:ack}

The authors would like to thank the Fundacion Valenciaport for their support and collaboration during the development and curation of the dataset.

% APPENDICES
\begin{appendices}

\section{Soroll-IA Sound Event Taxonomy}\label{app:taxonomy}

\begin{longtable}{p{3.2cm} p{9.2cm}}
\caption{Soroll-IA sound event taxonomy. Each class is weakly annotated at the clip level and reflects typical acoustic events in industrial port environments.}
\label{tab:soroll_classes} \\

\hline
\textbf{Class} & \textbf{Description} \\
\hline
\endfirsthead

\hline
\textbf{Class} & \textbf{Description} \\
\hline
\endhead

Air brake & Release or activation of pneumatic braking systems, typically associated with heavy vehicles or trains operating within the port. \\
Beep, bleep & Constant electronic warning tones emitted by vehicles or machinery, often used for alerts or signaling. \\
Bird vocalization, bird call, bird song & Vocal sounds produced by birds present in the port environment, including calls and songs. \\
Blower & Continuous or semi-continuous noise generated by industrial ventilation or air-moving equipment. \\
Clang & Sharp metallic impact sounds resulting from the collision of metal objects or structures. \\
Cluck & Short, impulsive clicking sounds produced by mechanical components or brief contact events. \\
Construction noises & Composite sounds associated with construction activities, including multiple overlapping tools and machinery. \\
Engine & Continuous noise produced by combustion engines of vehicles, machinery, or ships operating nearby. \\
Idling & Low-frequency, steady engine noise emitted while a vehicle or machine is running without active movement. \\
Jackhammer & Repetitive, high-energy percussive sounds generated by pneumatic or electric jackhammers. \\
Jingle, tinkle & Light, high-pitched metallic sounds caused by small objects or components vibrating or colliding. \\
Metal noises & General metallic sounds not covered by more specific classes, including scraping, resonances, or impacts. \\
Motor vehicle (road) & Sounds produced by road vehicles in motion, including cars, trucks, and vans driving near the sensor. \\
Motorcycle & Acoustic emissions from motorcycles, characterized by higher-pitched and more variable engine sounds. \\
Railroad car, train wagon & Rolling noise generated by train wagons moving along railway tracks within the port area. \\
Reversing beeps & Periodic warning beeps emitted by vehicles when reversing, commonly used for safety in industrial zones. \\
Scrape & Frictional sounds produced when objects slide or scrape against hard surfaces. \\
Seagull & Vocalizations produced by seagulls commonly found in coastal and port environments. \\
Siren & Warning sirens emitted by port cranes, vehicles, or safety systems to signal operational states or hazards. \\
Squeak & High-pitched, tonal sounds caused by friction, typically from moving mechanical parts. \\
Tick & Very short, impulsive clicking sounds, often produced by small mechanical or electrical components. \\
Tick-tock & Repetitive ticking sounds with a regular temporal pattern, resembling clock-like mechanisms. \\
Traffic noise, roadway noise & Aggregate noise from multiple road vehicles, reflecting continuous traffic activity near the port. \\
Train wheels squealing & High-pitched squealing sounds generated by train wheels during braking or tight curve negotiation. \\
Truck & Sounds generated by truck trailers, including structural vibrations, cargo movement, coupling noises, and rolling-related acoustic emissions, excluding engine-related sounds. \\
Wind & Ambient noise caused by wind interacting with the environment and the recording hardware. \\

\hline
\end{longtable}

%%=============================================%%
%% For submissions to Nature Portfolio Journals %%
%% please use the heading ``Extended Data''.   %%
%%=============================================%%

%%=============================================================%%
%% Sample for another appendix section			       %%
%%=============================================================%%

%% \section{Example of another appendix section}\label{secA2}%
%% Appendices may be used for helpful, supporting or essential material that would otherwise 
%% clutter, break up or be distracting to the text. Appendices can consist of sections, figures, 
%% tables and equations etc.

\end{appendices}

%%===========================================================================================%%
%% If you are submitting to one of the Nature Portfolio journals, using the eJP submission   %%
%% system, please include the references within the manuscript file itself. You may do this  %%
%% by copying the reference list from your .bbl file, paste it into the main manuscript .tex %%
%% file, and delete the associated \verb+\bibliography+ commands.                            %%
%%===========================================================================================%%
\FloatBarrier
\bibliographystyle{unsrt} % Citas numéricas ordenadas por aparición
\bibliography{sn-bibliography} % Sin la extensión .bib

@inproceedings{urbansound8k,
  title={A dataset and taxonomy for urban sound research},
  author={Salamon, Justin and Jacoby, Christopher and Bello, Juan Pablo},
  booktitle={Proceedings of the 22nd ACM international conference on Multimedia},
  pages={1041--1044},
  year={2014}
}

@article{triantafyllopoulos2025computer,
  title={Computer audition: From task-specific machine learning to foundation models},
  author={Triantafyllopoulos, Andreas and Tsangko, Iosif and Gebhard, Alexander and Mesaros, Annamaria and Virtanen, Tuomas and Schuller, Bj{\"o}rn W},
  journal={Proceedings of the IEEE},
  year={2025},
  publisher={IEEE}
}

@article{gong2021psla,
  title={Psla: Improving audio tagging with pretraining, sampling, labeling, and aggregation},
  author={Gong, Yuan and Chung, Yu-An and Glass, James},
  journal={IEEE/ACM Transactions on Audio, Speech, and Language Processing},
  volume={29},
  pages={3292--3306},
  year={2021},
  publisher={IEEE}
}

@article{singh2024atgnn,
  title={Atgnn: Audio tagging graph neural network},
  author={Singh, Shubhr and Steinmetz, Christian J and Benetos, Emmanouil and Phan, Huy and Stowell, Dan},
  journal={IEEE Signal Processing Letters},
  volume={31},
  pages={825--829},
  year={2024},
  publisher={IEEE}
}

@inproceedings{schmid2023efficient,
  title={Efficient large-scale audio tagging via transformer-to-cnn knowledge distillation},
  author={Schmid, Florian and Koutini, Khaled and Widmer, Gerhard},
  booktitle={ICASSP 2023-2023 IEEE international Conference on acoustics, Speech and signal processing (ICASSP)},
  pages={1--5},
  year={2023},
  organization={IEEE}
}

@article{mesaros2021sound,
  title={Sound event detection: A tutorial},
  author={Mesaros, Annamaria and Heittola, Toni and Virtanen, Tuomas and Plumbley, Mark D},
  journal={IEEE Signal Processing Magazine},
  volume={38},
  number={5},
  pages={67--83},
  year={2021},
  publisher={IEEE}
}

@inproceedings{schmid2025effective,
  title={Effective pre-training of audio transformers for sound event detection},
  author={Schmid, Florian and Morocutti, Tobias and Foscarin, Francesco and Schl{\"u}ter, Jan and Primus, Paul and Widmer, Gerhard},
  booktitle={ICASSP 2025-2025 IEEE International Conference on Acoustics, Speech and Signal Processing (ICASSP)},
  pages={1--5},
  year={2025},
  organization={IEEE}
}

@inproceedings{gemmeke2017audio,
  title={Audio set: An ontology and human-labeled dataset for audio events},
  author={Gemmeke, Jort F and Ellis, Daniel PW and Freedman, Dylan and Jansen, Aren and Lawrence, Wade and Moore, R Channing and Plakal, Manoj and Ritter, Marvin},
  booktitle={2017 IEEE international conference on acoustics, speech and signal processing (ICASSP)},
  pages={776--780},
  year={2017},
  organization={IEEE}
}

@article{fonseca2021fsd50k,
  title={Fsd50k: an open dataset of human-labeled sound events},
  author={Fonseca, Eduardo and Favory, Xavier and Pons, Jordi and Font, Frederic and Serra, Xavier},
  journal={IEEE/ACM Transactions on Audio, Speech, and Language Processing},
  volume={30},
  pages={829--852},
  year={2021},
  publisher={IEEE}
}

@inproceedings{Purohit2019,
    author = "Purohit, Harsh and Tanabe, Ryo and Ichige, Takeshi and Endo, Takashi and Nikaido, Yuki and Suefusa, Kaori and Kawaguchi, Yohei",
    title = "MIMII Dataset: Sound Dataset for Malfunctioning Industrial Machine Investigation and Inspection",
    booktitle = "Proceedings of the Detection and Classification of Acoustic Scenes and Events 2019 Workshop (DCASE2019)",
    address = "New York University, NY, USA",
    month = "October",
    year = "2019",
    pages = "209--213"
}

@inproceedings{Zinemanas2019,
    author = "Zinemanas, Pablo and Cancela, Pablo and Rocamora, Martín",
    title = "MAVD: A Dataset for Sound Event Detection in Urban Environments",
    booktitle = "Proceedings of the Detection and Classification of Acoustic Scenes and Events 2019 Workshop (DCASE2019)",
    address = "New York University, NY, USA",
    month = "October",
    year = "2019",
    pages = "263--267"
}

@INPROCEEDINGS{mimiidue,
  author={Tanabe, Ryo and Purohit, Harsh and Dohi, Kota and Endo, Takashi and Nikaido, Yuki and Nakamura, Toshiki and Kawaguchi, Yohei},
  booktitle={2021 IEEE Workshop on Applications of Signal Processing to Audio and Acoustics (WASPAA)}, 
  title={MIMII Due: Sound Dataset for Malfunctioning Industrial Machine Investigation and Inspection with Domain Shifts Due to Changes in Operational and Environmental Conditions}, 
  year={2021},
  volume={},
  number={},
  pages={21-25},
  doi={10.1109/WASPAA52581.2021.9632802}}

@inproceedings{abesser2021idmt,
  title={IDMT-traffic: an open benchmark dataset for acoustic traffic monitoring research},
  author={Abe{\ss}er, Jakob and Gourishetti, Saichand and K{\'a}tai, Andr{\'a}s and Clau{\ss}, Tobias and Sharma, Prachi and Liebetrau, Judith},
  booktitle={2021 29th European Signal Processing Conference (EUSIPCO)},
  pages={551--555},
  year={2021},
  organization={IEEE}
}

@inproceedings{Heittola2020,
    author = "Heittola, Toni and Mesaros, Annamaria and Virtanen, Tuomas",
    title = "Acoustic scene classification in DCASE 2020 Challenge: generalization across devices and low complexity solutions",
    booktitle = "Proceedings of the Detection and Classification of Acoustic Scenes and Events 2020 Workshop (DCASE2020)",
    year = "2020",
    pages = "56--60",
    url = "https://arxiv.org/abs/2005.14623"
}

@inproceedings{Mesaros2018_DCASE,
    Author = "Mesaros, Annamaria and Heittola, Toni and Virtanen, Tuomas",
    title = "A multi-device dataset for urban acoustic scene classification",
    year = "2018",
    booktitle = "Proceedings of the Detection and Classification of Acoustic Scenes and Events 2018 Workshop (DCASE2018)",
    month = "November",
    pages = "9--13",
    url = "https://arxiv.org/abs/1807.09840"
}

@inproceedings{Naranjo:2025:AES,
  author    = {J. Naranjo-Alcazar and J. Grau-Haro and R. Ribes-Serrano and P. Zuccarello},
  title     = {Real-Time Audio Monitoring Pipeline with Edge Inference and IoT Supervision via ThingsBoard},
  booktitle = {AES International Conference on Machine Learning and Artificial Intelligence for Audio},
  publisher = {Audio Engineering Society},
  month     = {september},
  year      = {2025},
  address   = {London, UK},
  url       = {https://drive.google.com/file/d/10_oAOetJyphLBiMC8b4zZQzodFn-BIJm/view},
  note      = {Late Breaking Demo Paper}
}

@InProceedings{aes_berlin,
author="Naranjo-Alcazar, Javier
and Grau-Haro, Jordi
and Ribes-Serrano, Ruben
and Zuccarello, Pedro",
editor="Arai, Kohei",
title="A Data-Centric Framework for Machine Listening Projects: Addressing Large-Scale Data Acquisition and Labeling Through Active Learning",
booktitle="Advances in Information and Communication",
year="2025",
publisher="Springer Nature Switzerland",
address="Cham",
pages="647--659",
isbn="978-3-031-84457-7"
}

@inproceedings{khandelwal2023leveraging,
  title={Leveraging audio-tagging assisted sound event detection using weakified strong labels and frequency dynamic convolutions},
  author={Khandelwal, Tanmay and Das, Rohan Kumar and Koh, Andrew and Chng, Eng Siong},
  booktitle={2023 IEEE Statistical Signal Processing Workshop (SSP)},
  pages={329--333},
  year={2023},
  organization={IEEE}
}

@inproceedings{koh2022automated,
  title={Automated audio captioning using transfer learning and reconstruction latent space similarity regularization},
  author={Koh, Andrew and Fuzhao, Xue and Siong, Chng Eng},
  booktitle={ICASSP 2022-2022 IEEE International Conference on Acoustics, Speech and Signal Processing (ICASSP)},
  pages={7722--7726},
  year={2022},
  organization={IEEE}
}

@inproceedings{liu2022leveraging,
  title={Leveraging pre-trained bert for audio captioning},
  author={Liu, Xubo and Mei, Xinhao and Huang, Qiushi and Sun, Jianyuan and Zhao, Jinzheng and Liu, Haohe and Plumbley, Mark D and Kilic, Volkan and Wang, Wenwu},
  booktitle={2022 30th European Signal Processing Conference (EUSIPCO)},
  pages={1145--1149},
  year={2022},
  organization={IEEE}
}

@article{xiao2023graph,
  title={Graph attention for automated audio captioning},
  author={Xiao, Feiyang and Guan, Jian and Zhu, Qiaoxi and Wang, Wenwu},
  journal={IEEE signal processing letters},
  volume={30},
  pages={413--417},
  year={2023},
  publisher={IEEE}
}

@inproceedings{piczak2015esc,
  title={ESC: Dataset for environmental sound classification},
  author={Piczak, Karol J},
  booktitle={Proceedings of the 23rd ACM international conference on Multimedia},
  pages={1015--1018},
  year={2015}
}

@INPROCEEDINGS{chime,
  author={Foster, Peter and Sigtia, Siddharth and Krstulovic, Sacha and Barker, Jon and Plumbley, Mark D.},
  booktitle={2015 IEEE Workshop on Applications of Signal Processing to Audio and Acoustics (WASPAA)}, 
  title={Chime-home: A dataset for sound source recognition in a domestic environment}, 
  year={2015},
  volume={},
  number={},
  pages={1-5},
  doi={10.1109/WASPAA.2015.7336899}}

@article{bello2019sonyc,
  title={Sonyc: A system for monitoring, analyzing, and mitigating urban noise pollution},
  author={Bello, Juan P and Silva, Claudio and Nov, Oded and Dubois, R Luke and Arora, Anish and Salamon, Justin and Mydlarz, Charles and Doraiswamy, Harish},
  journal={Communications of the ACM},
  volume={62},
  number={2},
  pages={68--77},
  year={2019},
  publisher={ACM New York, NY, USA}
}

@Article{informatics11040071,
AUTHOR = {Andriulo, Francesco Cosimo and Fiore, Marco and Mongiello, Marina and Traversa, Emanuele and Zizzo, Vera},
TITLE = {Edge Computing and Cloud Computing for Internet of Things: A Review},
JOURNAL = {Informatics},
VOLUME = {11},
YEAR = {2024},
NUMBER = {4},
ARTICLE-NUMBER = {71},
URL = {https://www.mdpi.com/2227-9709/11/4/71},
ISSN = {2227-9709},
DOI = {10.3390/informatics11040071}
}

@inproceedings{ooi2021strongly,
  title={A strongly-labelled polyphonic dataset of urban sounds with spatiotemporal context},
  author={Ooi, Kenneth and Watcharasupat, Karn N and Peksi, Santi and Karnapi, Furi Andi and Ong, Zhen-Ting and Chua, Danny and Leow, Hui-Wen and Kwok, Li-Long and Ng, Xin-Lei and Loh, Zhen-Ann and others},
  booktitle={2021 Asia-Pacific Signal and Information Processing Association Annual Summit and Conference (APSIPA ASC)},
  pages={982--988},
  year={2021},
  organization={IEEE}
}

@inproceedings{annotation,
author = {Cartwright, Mark and Dove, Graham and M\'{e}ndez M\'{e}ndez, Ana Elisa and Bello, Juan P. and Nov, Oded},
title = {Crowdsourcing Multi-label Audio Annotation Tasks with Citizen Scientists},
year = {2019},
isbn = {9781450359702},
publisher = {Association for Computing Machinery},
address = {New York, NY, USA},
url = {https://doi.org/10.1145/3290605.3300522},
doi = {10.1145/3290605.3300522},
booktitle = {Proceedings of the 2019 CHI Conference on Human Factors in Computing Systems},
pages = {1–11},
numpages = {11},
location = {Glasgow, Scotland Uk},
series = {CHI '19}
}

@inproceedings{Garcia-Ballesteros2026confidently, 
title={{Confidently Wrong: Evaluating AudioSet-Trained Models Under Real-World Deployment}}, 
author={Garcia-Ballesteros, Marta and Naranjo-Alcazar, Javier and Grau-Haro, Jordi and Ribes-Serrano, Ruben and Zuccarello, Pedro}, 
year={2026}, 
month={jun}, 
booktitle={Journal of the Audio Engineering Society}, 
publisher={}, 
number={10274}, 
organization={AES}, 
}
%% if required, the content of .bbl file can be included here once bbl is generated
%%\input sn-article.bbl

\end{document}